\documentclass{IEEEtran}
\usepackage{amsmath}		
\usepackage{amssymb}
\usepackage{graphicx}		
\usepackage{enumerate}		
\usepackage{hyperref}		
\usepackage{subcaption}     
\usepackage[all]{hypcap}	
\usepackage{relsize}		
\usepackage{caption}  

\usepackage{indentfirst}
\usepackage{array}          
\usepackage[margin=1in, paperwidth=8.5in, paperheight=11in]{geometry} 
\usepackage{booktabs}
\usepackage[section]{placeins}
\usepackage{todonotes}
\usepackage{listings}
\usepackage{color}
\usepackage{siunitx}
\usepackage{mathtools}
\usepackage{comment}
\usepackage{xcolor}
\usepackage{cite}
\usepackage{mathtools}

\DeclarePairedDelimiter{\floor}{\lfloor}{\rfloor}

\DeclarePairedDelimiterX{\infdivx}[2]{(}{)}{%
  #1\;\delimsize\|\;#2%
}

\newcommand{\brP}{{P}}

\newcommand{\var}[1]{\mbox{var}\left\{ #1 \right \} }

 \newcommand{\change}[1]{{\color{black} #1}}

\lstdefinestyle{mystyle}{
    backgroundcolor=\color{backcolour},   
    commentstyle=\color{codegreen},
    keywordstyle=\color{magenta},
    numberstyle=\tiny\color{codegray},
    stringstyle=\color{codepurple},
    basicstyle=\footnotesize,
    breakatwhitespace=false,         
    breaklines=true,                 
    captionpos=b,                    
    keepspaces=true,                 
    numbers=left,                    
    numbersep=5pt,                  
    showspaces=false,                
    showstringspaces=false,
    showtabs=false,                  
    tabsize=2
}
 
\newtheorem{theorem}{Theorem}
\newtheorem{corollary}{Corollary}

\newtheorem{lemma}{Lemma}

\title{Asymptotic Analysis of the Downlink in Cooperative Massive MIMO Systems}
\author{Itsik Bergel\thanks{\noindent Itsik Bergel (itsik.bergel@biu.ac.il) is with the Faculty of Engineering, Bar Ilan University, Ramat-Gan, Israel.}, Siddhartan Govindasamy \thanks{ Siddhartan Govindasamy (govindsi@bc.edu) is with the Department of Engineering, Boston College, Chestnut Hill, MA, USA. }}

\date{October 2022}

\begin{document}

\maketitle

    \begin{abstract}
        We consider the downlink of a \change{cooperative cellular} communications system, where several base-stations around each mobile cooperate and perform zero-forcing to reduce the received interference \change{at the mobile.} We derive closed-form expressions for the asymptotic performance of the network as the number of antennas per base station grows large. These expressions capture the trade off between various system parameters, and characterize the joint effect of noise and interference (where either noise or interference is asymptotically dominant and where both are asymptotically relevant).  The asymptotic results are verified using Monte Carlo simulations, which indicate that they are useful even when the number of antennas per base station is only moderately large. Additionally, we show that when the number of antennas per base station grows large,  power allocation can be optimized locally at each base station. We hence present a power allocation algorithm that achieves near optimal performance while significantly reducing the coordination overhead between base stations.  The presented analysis is significantly more challenging \change{than} the uplink analysis, due to the dependence between beamforming vectors of nearby base stations. This statistical dependence is handled by introducing novel bounds on marked shot-noise point processes with dependent marks, which are also useful in other contexts.
    \end{abstract}

\section{Introduction}

Multiple-input-multiple-output (MIMO) systems with large numbers of antennas,  
also known as massive MIMO, have received tremendous attention in the research community (e.g.,  \cite{marzetta2010noncooperative, rusek2012scaling,marzetta2015massive, larsson2014massive, bjornson2015massive, lu2014overview, marzetta2016fundamentals}). Early works on massive MIMO systems (e.g., \cite{marzetta2010noncooperative}) used maximal ratio-transmission (MRT) for the downlink or maximal ratio combining (MRC) for the uplink, to ensure that signals add constructively. Interference mitigation was accomplished by the approximate (and asymptotically exact) orthogonality of long channel vectors with random, independent fading coefficients.
 
 While MRT/MRC are optimal when the number of antennas at the base stations (BSs) go to infinity, for finite (and even large) numbers of antennas, significantly higher spectral efficiencies are achievable using other linear processing techniques. On the uplink, minimum-mean-square-error (MMSE) processing is the optimal linear processing scheme for receive beamforming  \cite{ bjornson2020making,Govindasamy2018Uplink, hoydis2013massive}. In contrast, in the downlink the precoding decisions of each BS affect  mobiles of all BSs, and optimal processing can only be achieved by numeric optimization.  
 
 Since  no optimal processing scheme is known for the downlink, several heuristic beamforming approaches were considered for interference mitigation. Most of these are variations of zero-forcing precoding, where  transmit beamformers are designed to produce (near) zero interference at certain mobiles  (e.g., \cite{yang2013performance, hoydis2013massive, jin2015massive,bjornson2016massive,ammar2020statistical,kountouris2009transmission}). In \cite{hoydis2013massive}, regularized zero-forcing was found to perform as well as MRT,  but with an order of magnitude fewer antennas. Additionally, \cite{bjornson2016massive} showed significantly larger spectral efficiencies with zero forcing compared to MRT, even when the number of antennas is $\sim 200$. Hence, although MRT is asymptotically optimal on the downlink, zero-forcing precoding is attractive even when the number of antennas per BS is quite large. 
 
\par

\change{In high-density networks, signals from other base stations may cause significant interference. Thus, we focus on BS cooperation, which promises significant performance improvement. In particular, we focus on user-centric cooperation, where all BSs up to a certain distance from each mobile protect the mobile from interference. This network structure is strongly related to cell-free systems (e.g., \cite{bjornson2020making,wang2020uplink,interdonato2020local}) which have received a lot of attention in the literature. Thus our results are also relevant to some cell-free scenarios. Cell-free systems may differ from our work in 3 main aspects: They typically use access points with a small number of antennas; They typically use joint processing of all BSs; and they cooperate both for interference mitigation and for signal transmission (joint signal transmission is not considered herein as its contribution is smaller, and it complicates the analysis).}

So far, all analysis of the downlink in \change{cooperative} massive MIMO networks was done either for specific BS locations, or through simulation (e.g., \cite{hoydis2013massive, bjornson2020making}). \change{As such, obtaining generalizable conclusions using these approaches is challenging, and requires extensive numerical simulations. On the other hand, in the approach taken in this work, we explicitly model how the BSs and mobiles are positioned in space, which enables us to obtain insights on system performance such as the impact of BS density and mobile distribution on the spectral efficiency. Such an approach can enable system designers to trade-off various parameters such as BS and mobile densities, number of antennas, number of cooperating base stations and spectral efficiency.} These approaches have been observed to  “lead to remarkably precise characterizations” \cite{george2017ergodic}. Further,  analysis of cellular networks that explicitly considers the BS and mobile distribution in space is quite challenging, even for single antenna systems \cite{dhillon2012modeling}.   Hence, an analysis of cooperative cellular networks with multiple antennas, where the spatial distribution of BSs and mobiles is explicitly considered, is useful and challenging.

Here, we derive closed-form expressions for the asymptotic behavior of the Signal-to-Interference-plus-Noise Ratio (SINR), which give insights on the network performance and allow system optimization. The analysis explicitly addresses the spatial distribution of BSs and mobiles. 

We also introduce a novel approach for asymptotic analysis that captures the effect of both noise and interference. This approach is important because asymptotic expressions are often used to generate performance  approximations for finite systems. Finite systems are affected by both interference and noise, while traditional asymptotic analysis either reaches a noise-limited regime or an interference-limited regime.

Additionally, 
our results indicate that with large number of antennas, the effect of each BS on its neighbors can be captured through the network's average transmitted
power per BS. This finding allows
us to present a simple and efficient power allocation algorithm for large networks which only
involves a minor coupling between the local processing done at each base station.

\change{To facilitate analysis, this paper considers a relatively simple channel model with perfect channel knowledge and independent fading. Yet, it is important to note that practical channels are often characterized both by fading correlations and estimation errors. E.g., taking advantage of the fading correlation was shown to reduce the channel estimation load and allow lower feedback rates in \cite{AnsumanAdhikary_2013,Yin_2013}.
 Furthermore,  \cite{Bjornson_2017}  showed that the throughput in cell-free MIMO is asymptotically unbounded in a correlated fading model that satisfies specific conditions. Yet, such models are significantly more complicated and so far, have not yielded closed-form expressions of the achievable data rates. In contrast, in this work, we provide expressions for the spectral efficiency that can be evaluated directly, without having to resort to simulations of a large number of network realizations. 
 
 Note that channel estimation for large networks is an active research area with significant recent progress (e.g., \cite{ghazanfari2021model, banerjee2022downlink} which use Bayesian and machine-learning approaches). As channel estimation is constantly improving, it is important to also understand the achievable system performance if channel estimation errors become negligible.}
\par
\change{We emphasize that the downlink scenario is especially challenging for analysis due to statistical dependence between the actions of the BSs which are coupled through the locations of nearby mobiles. This form of dependence is different from the correlation between the fading coefficients as considered in works such as \cite{wang2022uplink, wang2020uplink}. Thus, this work is the first to present asymptotic results for the performance of the downlink in a cooperative network. Note that several works have presented large network analysis but without closed-form expressions (e.g.\cite{bjornson2020making,interdonato2020local}). Such works still require Monte-Carlo simulations to obtain insight into system performance.}

Analysis of spatially distributed systems with statistical dependence \change{induced through spatial locations} is usually very complicated. Thus, this work was only enabled through the derivation of novel bounds on the second joint moments of shot noises driven by  marked Poisson Point processes (PPPs), where the marks are allowed to be statistically dependent\footnote{An introduction to shot noise processes and marked PPPs is given for example in \cite{Stoyan} and \cite{haenggi2012stochastic}.}. These novel bounds are especially useful for scenarios with weak dependence between the marks, where the bounds become tight. 
%
%
In this work we use these bounds
 to derive tight upper and lower bounds on the  joint moments of the interference and hence  to prove the  limited effect of the statistical dependence between the transmitting base stations. 

To emphasize the importance of our findings, we also present numerical results  based on Monte-Carlo simulations. These results show  agreement with the theoretical predictions. Moreover, they show that the theoretical predictions are valuable even when the numbers of antennas per BS are only moderately large.

\textcolor{black}{
In summary, the main contributions of this work are:
\begin{itemize}
    \item Closed-form expressions for the asymptotic performance in the downlink of cooperative cellular networks. 
    \item A novel approach for asymptotic analysis of the downlink of cooperative cellular networks that captures the effect of both noise and interference.
    \item A proof that the asymptotic effect of each BS on its nearby mobiles depends only on the network's average transmitted power per BS. We used this finding to derive a novel, low-complexity power allocation scheme for large networks. 
\item Novel bounds on the second joint moments of shot noises driven by marked Poisson Point processes (PPPs), where the marks are allowed to be statistically dependent.
\end{itemize}}

    \section{Problem Statement}
    \subsection{Network and channel model}
    We consider the downlink in a cellular network. The network transmission is done through BSs which are distributed according to a homogeneous Poisson Point Process (HPPP), $\Phi_b$, with density $\lambda_{\rm b}$ BSs per unit area. Each BSs has $L$ antennas and serves $M$ active mobiles, which are located at a distance of at most $R$ from the BS. Hence, the area density of mobiles is  $\lambda = \lambda_{\rm b} M$. Each mobile has one antenna.

    Without loss of generality, we analyze the performance of a \emph{typical} mobile located at the origin. The typical mobile is labeled as mobile-0, and is served by BS-0. 
     BSs and mobiles are indexed separately and we denote by $r_{i,k}$ the distance between mobile $i$ and BS $k$.
    
    Let  the BS that serves  mobile $i$ be $b_i$ (and hence $r_{i,b_i}\le R$). Also, let $\mathcal{M}_k = \left\{i: b_i = k \right\}$ be the set of mobiles served by BS $k$  (with $|\mathcal{M}_k|=M$). Note that given the existence of the typical mobile at the origin, its serving  BS must be within radius $R$ of the origin (i.e., $r_{0,0}\le R$). All other BSs still form an HPPP  with a density of $\lambda_{\rm b}$. 
    
    Let $v_i$ be the symbol transmitted by BS $b_i$ to mobile-$i$ (with $E[v_{i}]=0$ and  $E[|v_{i}|^2]=1$) using a normalized beamforming vector $\mathbf{w}_i$ (with $\|\mathbf{w}_{i}\|=1$).    The specific form for $\mathbf{w}_i$ is defined in Subsection \ref{sec:PZF}. 
    For a specific symbol, the sampled received signal at mobile-$j$ is
    \begin{align}\notag
        u_j =  a_{b_j}\varphi_{j,{b_j}}\mathbf{h}_{j,{b_j}}^\dagger\mathbf{w}_j v_j +  \sum_{i\ne j} a_{b_i}\varphi_{i,b_i} \mathbf{h}_{j,b_i}^\dagger\mathbf{w}_{i} v_{i}  + n_j
    \end{align}
  where $\mathbf{h}_{j,k}$ is the (complex conjugate of the) channel between BS $k$ and mobile $j$, $n_j\sim \mathcal{CN}(0,\sigma^2)$ is  the additive Gaussian noise. $\mathcal{CN}(0,\sigma^2)$ refers to a circularly symmetric, zero mean,  complex Gaussian random variable with variance  $\sigma^2$.  
  The random variables $a_k \in \{0,1\}$ determine whether or not  BS $k$ is active, as detailed in Subsection \ref{sec:PZF}. $\varphi^2_{i,b_i}$ is the power allocated by BS $b_i$ to mobile $i$.    We enforce a maximum transmit power constraint on each BS, such that for each $k$ 
  $    \sum_{i\in\mathcal{M}_k} \varphi^2_{i,k} \le \brP
  $.
  The channel gain vector is given by:
  $      \mathbf{h}_{i,k}  = \mathbf{g}_{i,k} r_{i, k}^{-\frac{\alpha}{2}}
  $
  where  $\alpha > 2$ is the path loss exponent and $\mathbf{g}_{i,k}$ is the fading vector between mobile $i$ and BS $k$, which has i.i.d., $\mathcal{CN}(0,1)$ entries. 
 
  Thus, the SINR  for mobile $i$ is:
      \begin{align}\label{d:SINR}
        \eta_i =\frac{\left| a_{b_i}\varphi_{i,{b_i}}\mathbf{h}_{i,{b_i}}^\dagger\mathbf{w}_{i} \right|^2}{  \sum_{j \ne i }\left|a_{b_j} \varphi_{j,b_j} \mathbf{h}_{i,b_j}^\dagger\mathbf{w}_{j}\right|^2 + \sigma^2}.
    \end{align}
Assuming Gaussian codebooks are used to serve all mobiles, with long enough blocks  and system bandwidth $B$, the achievable rate for mobile $i$, denoted by $R_i$ is
\begin{IEEEeqnarray}{rCl}\label{d:R_i}
R_i&=&B\log_2(1+\eta_i)\,.
\end{IEEEeqnarray}

 \subsection{Partial Zero Forcing} \label{sec:PZF}
 
    The network establishes partial cooperation, where user data is not shared between BSs, but each BS obtains the channel state information (CSI) for all neighboring mobiles. 
This CSI is used by each  BS to perform zero-forcing on all mobiles within a radius $D$ of the BS.\footnote{ In some studies, all cooperating BS also share the user data and jointly transmit it to the user. Such schemes require a higher level of cooperation between the BSs and hence are not considered herein. We note that the analysis of such schemes can be done using the same approach (and has the same complexity) but requires a longer analysis.} 
Following our comment above, we take the approach of \cite{Zhu2018Stochastic} and focus on characterizing the spectral efficiencies that can be achieved if accurate channel estimation is possible.

As each BS has $L$ antennas, a proper choice of the precoding weight vector can null the signal at $L-1$ mobiles. 
If the number of mobiles in the disk of radius $D$  around a given BS exceeds $L$, then the BS cannot perform proper transmission and ZF. Moreover, if the distance to a served mobile is greater than $D$, the BS is capable of serving it if the number of users in the disk of radius $D$  does not exceed $L-1$.

In all cases, no BS interferes with a mobile at a distance of less than $D$. If the BS cannot serve a mobile without creating such interference, it has to adopt an alternative transmission strategy.  As the choice of active users is beyond the scope of this paper, we consider here a simple scheme, in which a BS transmits only if it can serve all of its users without violating the interference constraint. Thus,  a BS does not transmit at all if it has more than $L$ users within a radius $D$, or if it has $L$ users within this radius and at least one served mobile outside of this radius.
  Hence
\begin{align}
    a_k=\begin{cases}
    1 \text{ if } (\left|\mathcal{D}_k\right| < L)\, \mbox{or}\, (\left|\mathcal{D}_k\cup\mathcal{M}_k\right| \leq L),  \\
    0 \text{ otherwise. }  
    \end{cases} \label{eqn:BSActivationDefinition}
\end{align}
where  $\mathcal{D}_k=\{i:r_{i,k}<D\}$ is the set of mobiles in the radius $D$ disk around BS $k$.
 We will ultimately show that the probability of no transmission ($a_k = 0$) becomes negligible in the regime of interest, where $L$ is large.

While we assume that the CSI at  BS $k$ includes the channel vectors  $\mathbf{h}_{i,k}$  $\forall i \in \mathcal{D}_k \cup \mathcal{M}_k$, for analysis purposes, it is more convenient to use only the fading vectors, $\mathbf{g}_{i,k}$. Note that the resulting ZF precoder is identical whether the BSs use $\mathbf{g}_{i,k}$ or $\mathbf{h}_{i,k}$ to construct the precoders.  Thus, for  served user $i$ (with $b_i=k$) BS $k$ constructs the interference channel matrix, $\mathbf{G}_{\bar i,k}\in \mathbb{C}^{L\times (|\mathcal{D}_k\setminus \{i\}|) }$ whose columns are the vectors $\mathbf{g}_{\ell,k}$ for all $\ell\in\mathcal{D}_k\setminus \{i\}$. Then, the precoding weight vector used by BS $k$ for user $i$ is given by:
\begin{align}\label{d:w_ik}
    \mathbf{w}_{i, k} = 
     \frac{\mathbf{Q}_{i,k}\mathbf{g}_{i,k}}{\|\mathbf{Q}_{i,k}\mathbf{g}_{i,k}\|}
\end{align}
where $\mathbf{Q}_{i,k}=\mathbf I - \mathbf{G}_{\bar i,k}(\mathbf{G}_{\bar i,k}^H\mathbf{G}_{\bar i,k})^{-1}\mathbf{G}_{\bar i,k}^H$ is a projection matrix onto the null space of $ \mathbf{G}_{\bar i, k}$.

\section{Asymptotic behavior of the SINR}
In this section, we characterize the asymptotic limit of the typical mobile's SINR as the number of antennas increases. The analysis herein is enabled by the following novel bounds on shot noise functions over marked PPP. The uniqueness of these bounds is that they hold even if the marks are statistically dependent. The bounds are presented as a theorem because they are very general, and can be applied to various scenarios. 

\begin{theorem}[Bounds on the second joint moment of shot noise functions]\label{th:SN_bounds}
Let $\Phi$ be a marked PPP over $\mathbb R^2$ with density $\lambda(r,\theta)$, where each point $(r_i,\phi_i)$ is associated with two random marks $p_{i,1}$ and $p_{i,2}$. Also, let $I_1$ and $I_2$ be shot noise functions of the form  $I_k=\sum_{i} f_k(r_i)\cdot p_{i,k}$ for continuous functions $f_k(\cdot)$. If there exists three continuous functions $q_0(r),q_1(r),q_2(r)$ such that 
\begin{align}
    E\left[\left. p_{i,1}p_{j,2}\right| r_{i},r_{j}\right]\le (\ge) \begin{cases}
    q_1(r_{i})q_2(r_{j}) & i\ne j \\
  q_0(r_{i}) & i=j
  \end{cases}\label{Eqn:JointBounds}
\end{align}
is satisfied
for any two points with indexes $i$,$j$, then the second joint moment of the shot noises is lower (upper) bounded by:
\begin{align}
    E\left[I_1 I_2\right]\le (\ge)
    &\prod_{k=1}^2\left(2\pi \int_0^\infty f_k(r) q_k(r) \bar\lambda(r)r dr\right)+\notag\\
    &2\pi \int_0^\infty f_1(r) f_2(r) q_0(r) \bar\lambda(r)r dr 
\end{align}
where 
$\bar \lambda(r)=\frac{1}{2\pi}\int_0^{2\pi }\lambda(r,\theta)d\theta$.
\end{theorem}
{\it Proof: See Appendix \ref{Sec:ProofOfCrossExpectation}}

 Before we present our main result, we need to discuss the behavior of the  ZF radius, $D$, and of the noise variance $\sigma^2$. As described below the SINR limit is less interesting if these quantities are constant. Hence we allow both to be a function of $L$. 

For the ZF radius, we set $D=s\cdot L^\beta$ (which includes constant $D$ as a special case). We will show in Appendix \ref{sec:InterfPowerConv} that if $\beta>0.5$ or if $\beta=0.5$ and $s\ge  1/\sqrt{\pi\lambda_b M}$, then asymptotically, each BS will try to zero more users than it has antennas. Thus the network will have a zero asymptotic rate. Hence, the rest of the manuscript will focus on $\beta<0.5$ or $\beta=0.5$ and $s^2\pi\lambda_b M<1$.

Typical network analysis assumes that the noise variance does not depend on the number of antennas. Yet, such networks are asymptotically (as the number of antennas grows large) noise limited. Thus, asymptotic analysis with constant noise variance cannot capture the trade off between noise and interference. In this work, we allow the noise to change as $\sigma^2=\mu L^{-\zeta}$. Such a dependence allows us to derive asymptotic expressions that characterize the interference limited regime, the noise limited regime, and the joint effect of noise and interference. In Section \ref{Sec:NumericalSimulations} we demonstrate that the resulting asymptotic expressions capture the joint effect of interference and noise even for finite networks where the number of antennas is moderately large.  

Denote by $\bar P$  the average transmit power of the BSs in the network.  In the following, we analyze  systems when the BSs apply any power allocation as long as the correlation of the total BS power on other BSs is decreasing with distance. Practically, this would apply to any reasonable  power allocation scheme. We mathematically require that for any BS:
\begin{align}\label{e:average_p_vs_distance}
\bar P(r_{0,k})\triangleq E\left[ \sum_{i\in {\mathcal{M}}_k}\varphi_{i,k}^2|a_k=1,r_{0,k}  \right]=\bar P + \xi_{0,k}
\end{align}
and
for any two BSs, denoting their distance by $\tilde r_{k,\ell}$:
\begin{align}\notag
    &E\left[\sum_{i\in \mathcal{M}_k}\varphi_{i,k}^2 \sum_{j\in \mathcal{M}_\ell}\varphi_{j,\ell}^2\Big|\Phi_b,a_k=a_\ell=1\right] = \bar P^2 + \tilde\xi_{k,\ell}
\end{align}
where $|\xi_{0,k}|\le \delta r_{0, k}^{-\gamma}$ and $|\tilde \xi_{k,\ell}|\le \min\{P^2,\delta \tilde r_{k,\ell}^{-\gamma}\}$
  with $\gamma>0$.

With these assumptions, the next theorem describes the asymptotic limit of $\eta_0$, the SINR of the typical user:
\begin{theorem}\label{t:SINR_limit}
If $\zeta< \beta(\alpha-2)$, then
\begin{align}\label{eq:Theorem_Lim_1}
    \lim_{L\to\infty}L^{-1-\zeta}\eta_0  =  (1 - \tilde s^2\pi\lambda_b M)\frac{\varphi_{0,0}^2r^{-\alpha}_{0,0}}{\mu}  
\end{align}
where $\tilde s=s$ if $\beta=0.5$ and zero if $\beta<0.5$.
If $\zeta\ge \beta(\alpha-2)$, then 
\begin{align}\label{eq:Theorem_Lim_2}
    \lim_{L\to\infty} L^{-1-\beta(\alpha-2)}\eta_0 = \frac{(1 - \tilde s^2\pi\lambda_b M)\varphi_{0,0}^2r^{-\alpha}_{0,0}}{\frac{1}{s}\frac{2\pi\bar P\lambda_b}{\alpha-2} +\tilde \mu }
\end{align}
where $\tilde \mu=\mu$ if $\zeta=\beta(\alpha-2)$ and zero otherwise. Both limits hold in probability.
\end{theorem}
\noindent {\it Proof:} 
We show the convergence of appropriately scaled interference and signal powers at the typical mobile, and then combine them to complete the proof. 

Define the interference power at the typical mobile as 
\begin{align}
 I_0 &= \sum_{i = 1}^\infty\left|a_{b_i}\varphi_{i,b_i} \mathbf{h}_{0,b_i}^\dagger\mathbf{w}_{i}\right|^2
\end{align}
The convergence of $I_0$ with appropriate normalization is given by the following lemma:
\begin{lemma}\label{lemma:InterfConv}
For any $s>0$ with $0<\beta<0.5$ or $0<s<1$ with $\beta=0.5$, we have in mean-square error (MSE):
\begin{align}
    \lim_{D\to\infty}D^{\alpha-2}I_0 &= \frac{2\pi\bar P\lambda_b}{\alpha-2}, 
    \label{Eqn:BSFocusScaledUBLimitMS} 
\end{align}
and $\var{D^{\alpha-2}I_0} = O(D^{4-2\alpha}).$
\end{lemma}
\noindent {\it Proof:} See Appendix \ref{sec:InterfPowerConv}.

Next, we define the signal power at the typical mobile:
\begin{align}
    S_0&=\left| \varphi_{0,0} \, a_0 \, r_{0, 0}^{-\alpha/2} \, \mathbf{g}^\dagger_{0, 0} \, \frac{\mathbf{Q}_{0,0} \mathbf{g}_{0,0}}{\|\mathbf{Q}_{0,0}\mathbf{g}_{0,0}\|}\right|^2\label{eqn:S0def}
\end{align}
 The convergence of $S_0$ 
 is given by:

\begin{lemma}  \label{Lemma:SignalPower}
The normalized signal power converges in probability as follows:
\begin{align}
    &\lim_{L\to\infty}\frac{S_0}{L  - s^2L^{2\beta}\pi\lambda_b M}  
    =\varphi_{0,0}^2r^{-\alpha}_{0,0}  \label{Eqn:SignalPowerLimit}
\end{align}
\end{lemma}
{\it Proof: See Appendix \ref{Lemma:SignalPowerProof}}

Combining \eqref{Eqn:BSFocusScaledUBLimitMS} with the noise scaling, $\sigma^2=\mu L^{-\zeta}$, and the ZF radius scaling, $D=s\cdot L^\beta$, we see that
if $\zeta< \beta(\alpha-2)$, the noise plus interference satisfies:
\begin{align}\label{e:IplusN_small_zeta}
    \lim_{L\to\infty}L^{\zeta}(I_0 +\sigma^2)&=\mu 
\end{align}
while if $\zeta\ge \beta(\alpha-2)$ the noise plus interference satisfies:
\begin{align}\label{e:IplusN_large_zeta}
    \lim_{L\to\infty}L^{\beta(\alpha-2)}(I +\sigma^2)&= \frac{1}{s}\frac{2\pi\bar P\lambda_b}{\alpha-2} +\tilde \mu . 
\end{align}
in MSE (and hence, in probability) where $\tilde \mu=\mu$ if $\zeta=\beta(\alpha-2)$ and zero otherwise.  
Combining \eqref{e:IplusN_small_zeta}, \eqref{e:IplusN_large_zeta} and \eqref{Eqn:SignalPowerLimit} proves  convergence of \eqref{eq:Theorem_Lim_1} and \eqref{eq:Theorem_Lim_2} in probability. \hfill $\blacksquare$


The  asymptotic expressions of Theorem \ref{t:SINR_limit} are very useful and give important insights on the network behavior. They capture the effect of both noise and interference, and allow the evaluation of the optimal ZF radius. 
\textcolor{black}{It is important to note the different scenarios described by the values of $\zeta$ and $\beta$ in the theorem. Recall that $\zeta$ determines the scaling of the noise with the number of antennas. In the first scenario, if  $\zeta< \beta(\alpha-2)$ then the noise decreases too slowly and the network is noise-dominated (this also includes the case of $\zeta=0$). Thus, the performance in this case (12) are not affected by the interference. In the second scenario, if  $\zeta> \beta(\alpha-2)$ then the noise decreases too fast and the network is interference-dominated ($\tilde \mu=0$ in (13)). Thus, the most interesting scenario, where we see the effect of both noise and interference, is when $\zeta= \beta(\alpha-2)$. }

\textcolor{black}{As for the scaling of the ZF radius, $\beta$, we already showed that if $\beta>0.5$ then the ZF radius grows too fast and the network cannot handle the needed nulls. The theorem shows that if $\beta<0.5$, the ZF radius growth is slow enough such that it doesn't burden the network at all ($\tilde s=0$). Thus, optimal performance, which balances the effect of noise and interference, is achieved when $\beta=0.5$ and $\tilde s=s$. Further discussion on this optimal scenario, and on the achievable performance with a finite number of antennas are given in the numerical results section, below.}
Note that in the most interesting case, where $\beta=0.5$ (later shown to be optimal) and $\zeta=\alpha/2-1$, the limit simplifies to:
\begin{align}\label{eq:Theorem_Lim_3}
    \lim_{L\to\infty} L^{-\alpha/2}\eta_0 = \frac{(1 -  s^2\pi\lambda_b M)\varphi_{0,0}^2r^{-\alpha}_{0,0}}{\frac{1}{s}\frac{2\pi\bar P\lambda_b}{\alpha-2} + \mu }.
\end{align}
 This limiting expression for the SINR is particularly interesting as it captures the effect of both the interference and the noise and is hence useful for analyzing the SINR of practical systems with moderately large numbers of antennas (for which interference is still significant). The theorem also shows that the asymptotic SINR does not depend on the power allocation scheme, but only on the average power per BS ($\bar P$). This insight significantly simplifies the analysis of networks with different power allocation schemes. Further, it allows a simple derivation of an asymptotically  optimal power allocation scheme for massive MIMO networks, as shown in Section \ref{sec:PowerAllocation}.

\section{Optimization of the ZF radius}
Using Theorem \ref{t:SINR_limit} we can find the zero forcing radius, $D$, that maximizes the throughput, i.e., to find the optimal values of $\beta$ and $s$. If $\zeta<\beta(\alpha-2)$ the network is asymptotically noise limited. In this case, the ZF has no effect and optimization trivially results in $D=0$ (or in terms of \eqref{eq:Theorem_Lim_1}, $\tilde s=0$). Hence, from hereon, we focus only on the case that $\zeta\ge\beta(\alpha-2)$.

Inspecting \eqref{eq:Theorem_Lim_2}, the SINR, $\eta_0$ scales as $L^{1+\beta(\alpha-2)}$. Recalling that $\alpha>2$, the SINR is maximized when $\beta$ takes its maximal value, that is $\beta=0.5$. For $\beta = 0.5$, \eqref{eq:Theorem_Lim_3} shows that the SINR scales as $L^{\alpha/2}$. To optimize the SINR, we need to maximize the RHS of \eqref{eq:Theorem_Lim_3} with respect to $s$. 
Evaluating the derivative of the RHS of \eqref{eq:Theorem_Lim_3} with respect to $s$ and equating to zero, we 
conclude that the optimal value, $s^{*}$, must satisfy:
\begin{align} \label{eqn:Optimals}
   \frac{\alpha}{\alpha-2}\pi\lambda_b s^{*2} +  \frac{\mu}{\bar P} s^{*\alpha}=\frac{1}{M}.
\end{align}

While the optimal expression has no general closed-from solution, it can easily be solved numerically. For further insight, we can consider the no noise case ($\mu=0$) which leads to:
\begin{align}
s^*= \sqrt{\frac{\alpha - 2}{\alpha \pi \lambda_b M}}.
\end{align}
Thus, for large enough $L$, the best nulling radius is
\begin{align}
    D^* \approx \sqrt{\frac{\alpha - 2}{\alpha \pi \lambda_b M}L}. 
\end{align}
Since for large $L$ and $D$, the average number of mobiles in a radius $D$ disk is approximately $D^2 \pi  \lambda_b M$, on average, each BS protects approximately $(1 - 2/\alpha)L$ mobiles from interference.

\section{Throughput}

The characterization of the asymptotic SINR of each mobile in Theorem \ref{t:SINR_limit} is the main theoretic contribution of this manuscript. To better understand the implication of these results, it is beneficial to consider the total throughput of each BS in the network. For simplicity, this section focuses only on the more interesting case where $\beta=0.5$ and $\zeta=\beta(\alpha-2)=\alpha/2-1$.

The spectral efficiency of link $i$, $R_i$ is given by \eqref{d:R_i}. The following theorem uses \eqref{eq:Theorem_Lim_3} to characterize the asymptotic rate for user $i$ (recall that  mobile 0 is a typical mobile). It is important to note here that while Theorem \ref{t:SINR_limit} shows convergence in probability of the normalized SINR, Corollary \ref{T:MSE} shows convergence in MSE of the achievable rate. We need this stronger form of convergence in order to make assertions on the mean rate in the subsequent discussion. 
\begin{corollary} \label{T:MSE}
$R_i$ converges in MSE to:
\begin{align}\label{e:rate_convergence}
    \tilde R_i = \log_2\Big(1+L^{\alpha/2} \frac{(1  - s^2\pi \lambda_b M)\varphi^2_{i,b_i}r_{i,b_i}^{-\alpha}}{s^{2-\alpha}\frac{2\pi\bar P\lambda_b}{\alpha-2}+\mu}\Big).
\end{align}
That is,
$
\lim_{L\to\infty} E\Big[    \big|R_i - \tilde R_i\big|^2 \Big]= 0
$.
\end{corollary}
\begin{IEEEproof}
\eqref{e:rate_convergence} holds in probability by applying the continuous mapping theorem, in a manner similar to the proof of Theorem 2 in \cite{Govindasamy2018Uplink}. Proving convergence in MSE is more challenging and is given in Section \ref{Sec:MSEConvergence}.
\end{IEEEproof}

To further characterize the throughput, we need to limit the discussion and assume that the mobile locations admit a probability distribution (while so far we made no assumption on the mobiles except for their maximal distance from their BS). Additionally, as the network is ergodic,  we can characterize the network throughput by the throughput of the characteristic BS (BS $0$). Without loss of generality, we assume that  BS $0$ serves  mobiles $0,\ldots,M-1$. Thus, the average throughput per BS is:
\begin{align}
   R_{\textrm{av}}= E\left[\sum_{m=0}^{M-1}\log_2(1+\eta_{m})\right] .      
\end{align}

To make this derivation applicable also for finite SNRs, we define an asymptotic approximation of the average throughput per BS:
\begin{align}\label{d:Ras}
   R_{\textrm{as}}= E\left[\sum_{m=0}^{M-1}\log_2\left(1+
   \tilde R_m
   \right)\right].
\end{align}
We note that \eqref{e:rate_convergence} guarantees the convergence in MSE, hence
$
   \lim_{L\to \infty} \left(R-R_{\textrm{as}}\right)=0.
$
Thus, for large enough number of antennas, $L$, we can claim that $R\approx R_{\textrm{as}}$.

Fig 3. demonstrates the accuracy of the prediction in (22) for two different mobile distributions (mobiles on the cell edge, and uniformly randomly distributed within a cell). Markers show the simulated results while solid lines show the theoretical result based on \eqref{d:Ras}\footnote{Closed form expressions for the throughput in these scenarios can be easily derived, but are omitted here due to space constraints}. The cell radius is $R = 0.15$km, $\alpha = 3$, and the transmit power is set such that the SNR for a mobile at the cell edge is 10 dB from a single antenna.
\begin{figure}
    \centering
    \includegraphics[width=0.45\textwidth]{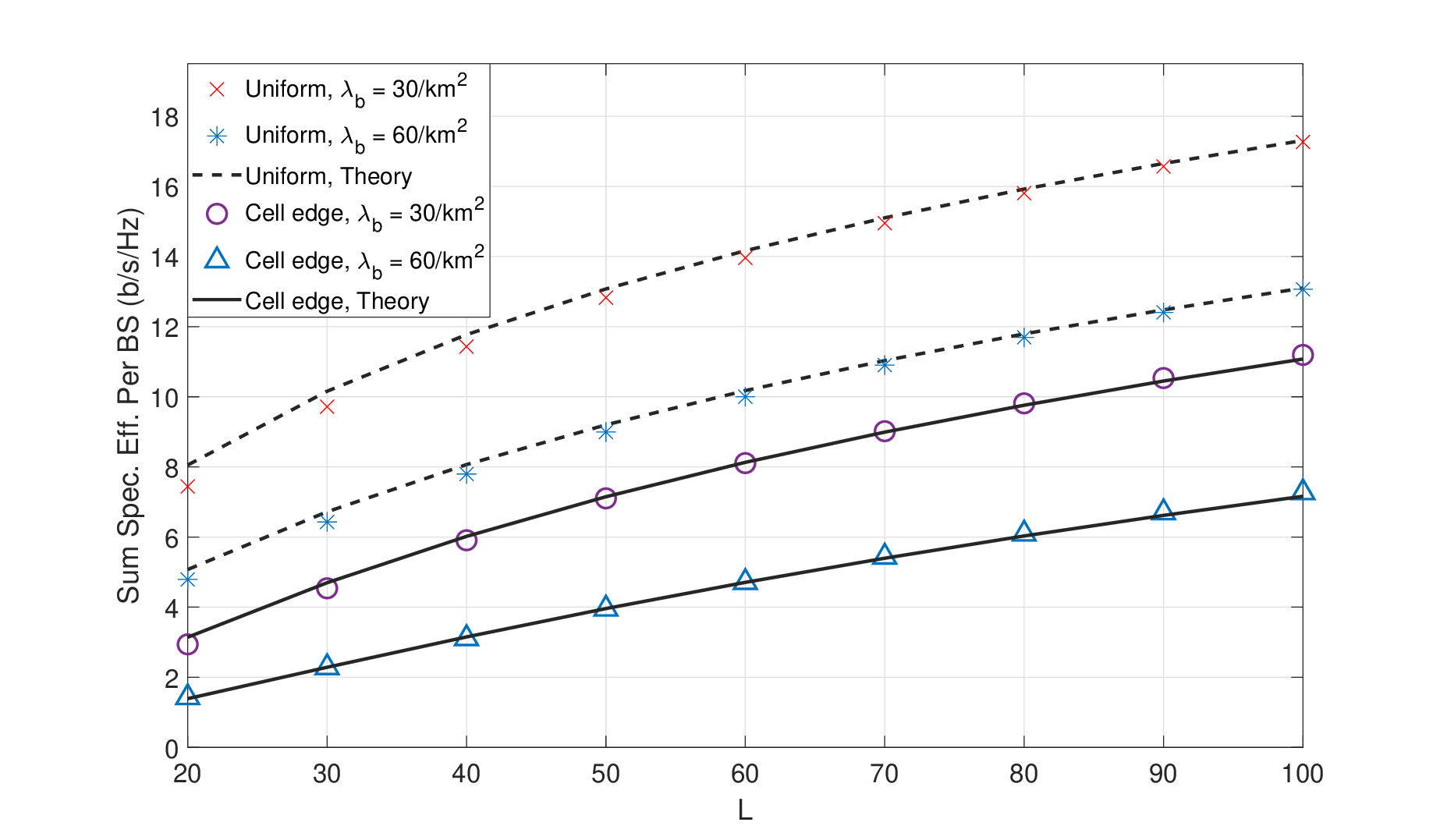}
    \caption{Spectral efficiency vs. number of antennas, $L$, for cell-edge and uniformly distributed mobiles. } 
    \label{fig:enter-label}
\end{figure}
\section{Power allocation}\label{sec:PowerAllocation}


In this section, we consider the power allocation to each mobile. We wish to  maximize 
\begin{align}\label{e:Ras2}
    R_{\textrm{as}}=\lambda_b E\left[\sum_{m=0}^{M-1}\log_2\left(1+g\varphi^2_{0,m}r_{0,m}^{-\alpha}\right)\right],
\end{align}
 subject to a peak power constraint $\sum_{m=0}^{M-1}\varphi^2_{0,m}\le P$, where $g= L^{\alpha/2}\frac{(1  - s^2\pi \lambda_b M)}{s^{2-\alpha}\frac{2\pi\bar P\lambda_b}{\alpha-2}+\mu}$.

That is, we search for $M$ functions $\varphi_{0,0}^2=f_0(\{r_{0,M}\}),\ldots,\varphi_{0,M-1}^2=f_{M-1}(\{r_{0,M}\})$ that maximize \eqref{e:Ras2} subject to the peak power constraint $0\le\sum_{m=0}^{M-1} f_m(\{r_{0,M}\})\le P$, where $\{r_{0,M}\}$ means $r_{0,0},\ldots,r_{0,M-1}$. Note that $g$ depends on the power allocation through the average transmission power $\bar P=E\left[\sum_{m=0}^{M-1}\varphi^2_{0,m}\right]$.
This is a calculus of variations problem (or an optimal control problem). 
Theorem \ref{T:water_filling} shows that it can be solved using a variation of the well known water filling algorithm. \begin{theorem}\label{T:water_filling}
The maximum rate of \eqref{e:Ras2}, subject to the peak power constraint, is given by:
\begin{align}\label{e:mx_pbar}
    \max_{0<\bar P\le P} \breve R (\bar P)
\end{align}
where $\breve R (\bar P)$ is the rate obtained using the power allocation functions $f_m(\{r_{0,M}\})$ evaluated by:
\begin{enumerate}
    \item For a given $\breve \lambda$, define 
    \begin{align}\label{e:Av_power_result}
        \breve f_m(\breve \lambda,\{r_{0,M}\})=\max\left\{0,\breve \lambda -\frac{1}{g r_{0,m}^{-\alpha}}\right\}
    \end{align}
    and 
    \begin{align}
        \hat P(\breve \lambda)=E\left[\min\left\{P,\sum_{m=0}^{M-1} \breve f_m(\breve \lambda,\{r_{0,M}\})\right\}\right]
    \end{align}
    \item Choose $\breve \lambda$ such that $\hat P(\breve \lambda)=\bar P$.
    \item If $\sum_{m=0}^{M-1} \breve f_m(\breve \lambda,\{r_{0,M}\})\le P$
    \begin{itemize}
        \item Set $f_m(\{r_{0,M}\})=\breve f_m(\breve \lambda,\{r_{0,M}\})$
    \end{itemize}
    else
    \begin{itemize}
        \item Set 
            \begin{align}\label{e:Pk_power_result}
         f_m(\{r_{0,M}\})=\max\left\{0,\breve \gamma -\frac{1}{g r_{0,m}^{-\alpha}}\right\}
    \end{align}
    where $\breve \gamma$ is chosen such that $\sum_{m=0}^{M-1}  f_m(\{r_{0,M}\})= P$.
    \end{itemize}
    \end{enumerate}
\end{theorem}

Furthermore, the optimization in \eqref{e:mx_pbar} is convex, and the choices of the constants $\breve \lambda$ and $\breve \gamma$ is over monotonic non-decreasing functions. Thus, the optimization problem can be solved  easily, using one dimensional convex optimization over a power allocation scheme with two water levels: a water level for the average BS power and a different water level for each BS that is constrained to the maximal peak power. Note that $\breve R (\bar P)$ is the maximal average BS throughput subject to an average power constraint $\bar P$.

\begin{IEEEproof}
We start by constraining the optimization in \eqref{e:Ras2} by the average BS power, $\bar P$. Since $g$ depends on $\bar P$, the additional constraint  significantly simplifies the problem (as $g$ becomes constant). The resulting optimization problem for a single BS given $\bar P$ is:
\begin{align}
    \breve R(\bar P)=\underset{f_0,\ldots,f_{M-1}}{\max}:&\ E\left[\sum_{m=0}^{M-1}\log_2\left(1+g f_m r_{0,m}^{-\alpha}\right)\right]
    \label{e:max_line}\\
    \text{Subject to}:&\ 
    \sum_{m=0}^{M-1} f_m\le P
    \label{c:peak_power}\\
    & \ E\left[\sum_{m=0}^{M-1} f_m\right]=\bar P
    \label{c:avg_power}
\end{align}
 where we have suppressed dependence on $r_{0,0},\cdots,r_{0,M-1}$ to simplify notation.

This is a convex maximization problem. The relevant variant of the Karush–Kuhn–Tucker (KKT) conditions (see for example Theorem 9.4 in \cite{clarke2013functional}) states that there exists scalar $\lambda$ and a function  $\gamma =\gamma(r_{0,0},\ldots,r_{0,M-1})\ge 0$ such that the  optimal solution must maximize 
\begin{align}\label{e:Lagrangian}
     \notag &E\left[\sum_{m=0}^{M-1}\log_2\left(1+g f_m r_{0,m}^{-\alpha}\right)\right]
    \\&+E\left[
    \gamma \sum_{m=0}^{M-1}  f_m\right ]
    +\lambda E\left[\sum_{m=0}^{M-1} f_m\right]
\end{align}
with respect to $f_m$ and must satisfy
\begin{align}\label{e:constraint}
E\left[
    \gamma \left(\sum_{m=0}^{M-1}  f_m -P\right)\right]=0
    \end{align}
and $E\left[\sum_{m=0}^{M-1}  f_m\right]=P.$
As we can choose $f_m$ for each set of distances, the maximization of  \eqref{e:Lagrangian} can be written as a set of maximizations:
\begin{align}\label{e:Lagrangian2}
     \max_{\phi_0,\ldots,\phi_{M-1}}& \sum_{m=0}^{M-1}\log_2\left(1+g \phi_m r_{0,m}^{-\alpha}\right)
    \notag\\&+
    \gamma \left(\sum_{m=0}^{M-1} \phi_m -P\right) 
    +\lambda \sum_{m=0}^{M-1} \phi_m
\end{align}
for each realization of the mobile distances. Note that \eqref{e:constraint} indicates that $\gamma\left(\sum_{m=0}^M \phi_m-P\right)= 0$ for any mobile distances (up to a set with zero probability). As long as we have a solution with $\sum_{m=0}^M \phi_m<P$ we set $\gamma=0$ and say that the constraint is not active. For these realizations we take the derivative of \eqref{e:Lagrangian2} with respect to $\phi_\ell $ and equate to zero:
\begin{align}
    0&= \frac{ g r_{0,\ell}^{-\alpha}\log_2 e}{1+g\phi_\ell r_{0,\ell}^{-\alpha}}
    +\lambda.
\end{align}
Simplifying and
defining also the water level
$\breve\lambda= -\frac{\log_2 e}{\lambda}$ leads to \eqref{e:Av_power_result}.
 
When \eqref{e:Av_power_result} leads to BS power that exceeds $P$, we have $\gamma>0$. In this case, 
 \eqref{e:constraint}  states the BS transmits at its max power, i.e., $\sum_{m=1}^M \phi_m =P$. This results in the simpler optimization 
 \begin{align}
     \max_{\phi_0,\ldots,\phi_{M-1}:\sum_{m=1}^M \phi_m -P} \sum_{m=1}^M\log_2\left(1+g \phi_m r_{0,m}^{-\alpha}\right)
\end{align}
which results in:
\begin{align}
      \varphi^2_\ell=-\frac{\log_2 e}{\gamma}
    -\frac{1}{g r_{0,m}^{-\alpha}}.
\end{align}
The secondary water level (associated with the peak power constraint for a specific realization) is then defined as
$\breve\gamma= -\frac{\log_2 e}{\lambda}$ , and leads to \eqref{e:Pk_power_result}.

The values of the water levels are determined to satisfy the average and peak power constraint (recalling that we have a scalar water level, $\breve\lambda$, for the average power constraint, and a separate water level, $\breve \gamma=\breve \gamma(r_{0,0},\ldots,r_{0,M-1})$, for each BS in which the peak power constraint is active. The monotonicity of the average and peak powers on their respective water level is obvious from the problem structure.
\end{IEEEproof}

\section{Numerical results} \label{Sec:NumericalSimulations}
In this section, we present numerical results that demonstrate our main assertions on the convergence of the normalized SINR and spectral efficiency, and show that localized power allocation (Theorem \ref{T:water_filling}) is sufficient even when the number of antennas per base station is only moderately large. For all simulations, the base stations and mobiles were placed on a square area with the edges wrapped around to reduce edge effects.  
\begin{figure}[t]
    \centering
    \includegraphics[width=0.5\textwidth]{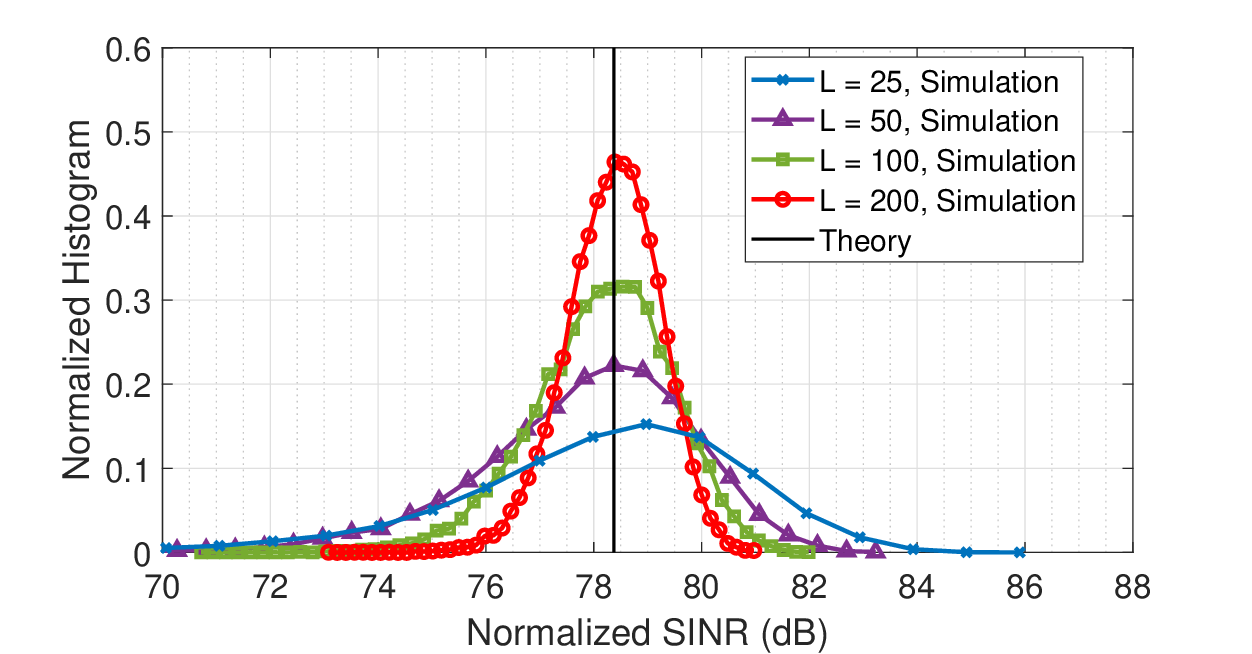}
    \caption{Histogram of the normalized SINR for fixed $\mu$. The vertical line shows the normalized SINR from \eqref{eqn:AsympNormSINR}.
    \label{fig:NormhistVarNoise}}
\end{figure}

\subsection{Normalized SINR}
To demonstrate the convergence of the signal and interference powers, first consider the normalized SINR of the mobiles in the network, where the SINR is normalized by $L^{\frac{\alpha}{2}}$, the transmit power and the path loss between each mobile and its serving base station. Thus, the normalized SINR for mobile $i$ is defined as
\begin{align}\label{d:normalized_SINR}
\underline \eta_i 
=\frac{1}{L^{\frac{\alpha}{2}}\phi_{i,b_i} r^{-\alpha}_{i, b_i} } \eta_i.
\end{align}

\begin{figure}[t]
    \centering
    \includegraphics[width=0.5\textwidth]{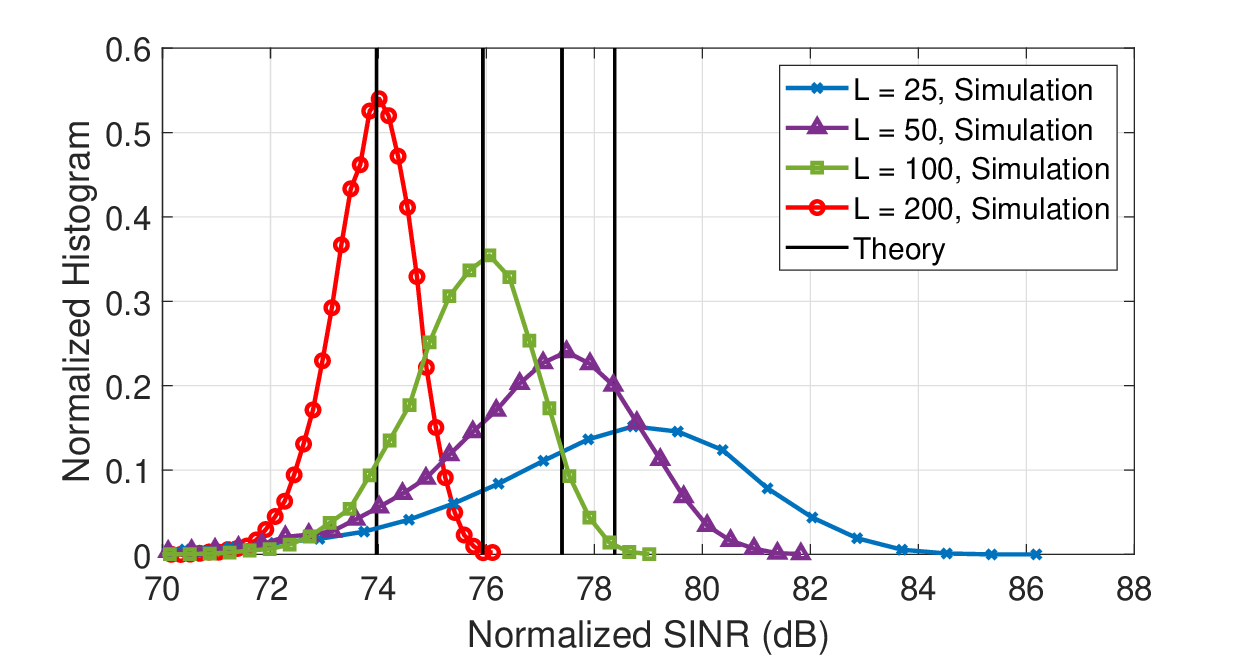}
    \caption{Histogram of the normalized SINR for fixed $\sigma^2$. The  vertical lines show the normalized SINR from \eqref{eqn:AsympApproxNormSINR}.}
    \label{fig:Normhist}
\end{figure}

Recall that the asymptotic analysis in the previous sections was made with an effective noise power that was changing with the number of antennas $L$. This is in contrast to the practical scenario where the noise power, $\sigma^2$, is fixed. To demonstrate this difference, Fig. \ref{fig:NormhistVarNoise} presents the  convergence of the interference power for the case of fixed $\mu$, while Fig. \ref{fig:Normhist} presents this convergence for the case of fixed noise power ($\sigma^2$). For the sake of practicality, all other  simulations (i.e., except those depicted in Fig. \ref{fig:NormhistVarNoise}) used the latter approach with fixed noise power. 
Strictly speaking, the asymptotic predictions of the previous section hold only for constant $\mu$ (i.e., when the noise variance changes with the number of antennas). In such case, from \eqref{eq:Theorem_Lim_3}:
\begin{align}
\lim_{L\to \infty}\underline \eta_i =  \frac{1  - s^2\pi \lambda_b M}{s^{2-\alpha}\frac{2\pi\bar P\lambda_b}{\alpha-2}+\mu}. \label{eqn:AsympNormSINR}
\end{align}

For the case of the normalized SINR, we simulated  networks of $1000$ base stations with density 30 BS/$km^2$, path loss exponnent of $\alpha=4$, $M=3$ active mobiles per base station and equal power allocated to each mobile.  Each configuration was simulated 10 times to generate sufficient data for the histograms.

Fig. \ref{fig:NormhistVarNoise} depicts  histograms of the  normalized SINRs, \eqref{d:normalized_SINR}, for a constant $\mu$ (and hence, noise power that decreases with $L$). The value of $\mu$ is  such that the SNR at the cell edge is 6dB for  $L =25$ (i.e., $25^{\alpha/2-1} P R^{-\alpha}/\mu =6$dB). The figure shows that even for $L = 25$, almost all the simulated normalized SINRs are within $\pm 6$dB of the asymptotic prediction given by the RHS of \eqref{eqn:AsympNormSINR}, which is indicated by the vertical  line. As the number of antennas increases, the normalized SINRs concentrate around the predicted value. For $L = 200$ antennas per base station, almost all the simulated normalized SINRs are within $\pm 2$dB of the predicted value. These results demonstrate that the normalized SINR does indeed converge to the asymptotic prediction. 

To derive theoretic predictions for the case of fixed noise power, we consider a finite number of antennas, and set $\mu = \sigma^2 L^{\frac{\alpha}{2}-1}$. Hence, $\mu$ becomes a function of $L$ instead of  being a constant. Yet, our asymptotic results show that for every $\mu$ there exists a large enough $L$ for which :
\begin{align}
\underline \eta_i \approx  \frac{1  - s^2\pi \lambda_b M}{s^{2-\alpha}\frac{2\pi\bar P\lambda_b}{\alpha-2}+\sigma^2 L^{\frac{\alpha}{2}-1}}. \label{eqn:AsympApproxNormSINR}
\end{align}
Simulations herein show that when $L$ is large enough, this approximation is very accurate, and gives good predictions even for the case of fixed noise power.

Fig. \ref{fig:Normhist} depicts the histograms of the normalized  SINRs  for a fixed noise power, $\sigma^2$. The value of $\sigma^2$ is set such that the SNR at the cell edge was 6dB (i.e., $P R^{-\alpha}/\sigma^2=6$dB). The theoretical predictions from \eqref{eqn:AsympApproxNormSINR} are shown as vertical lines. Note that in these simulations $\mu$ increases with $L$, and hence the asymptotic normalized SINR decreases with $L$. The convergence behavior is quite similar to the one depicted in Fig. \ref{fig:NormhistVarNoise}, with almost all of the simulated SINRs  within $\pm2$ dB of the theoretical prediction for $L = 200$. These simulations thus demonstrate that the asymptotic predictions are useful when the number of antennas is large, even when the noise power is constant. 


\subsection{Power Control}

As noted in Section \ref{sec:PowerAllocation}, the asymptotic results of this paper can be used to find the power allocation that maximizes the spectral efficiency. Further, most of the power allocation can be conducted on a per-base-station basis, resulting in a significant reduction in system complexity. 

To test the performance of the power allocation of Theorem \ref{T:water_filling} we compare its performance with two other policies, namely constant power allocation, where the transmit power for all mobiles is equal, and a power allocation policy based on the weighted-minimum-mean-square-error (WMMSE) sum-spectral efficiency maximizing algorithm from \cite{chakraborty2020efficient}. This latter algorithm is independent of the specific beamforming algorithm, and is applied here with the partial-zero-forcing approach.
Note that the complexity of the latter algorithm is very high for large networks.

\begin{figure}[t]
    \centering
    \includegraphics[width=0.5\textwidth]{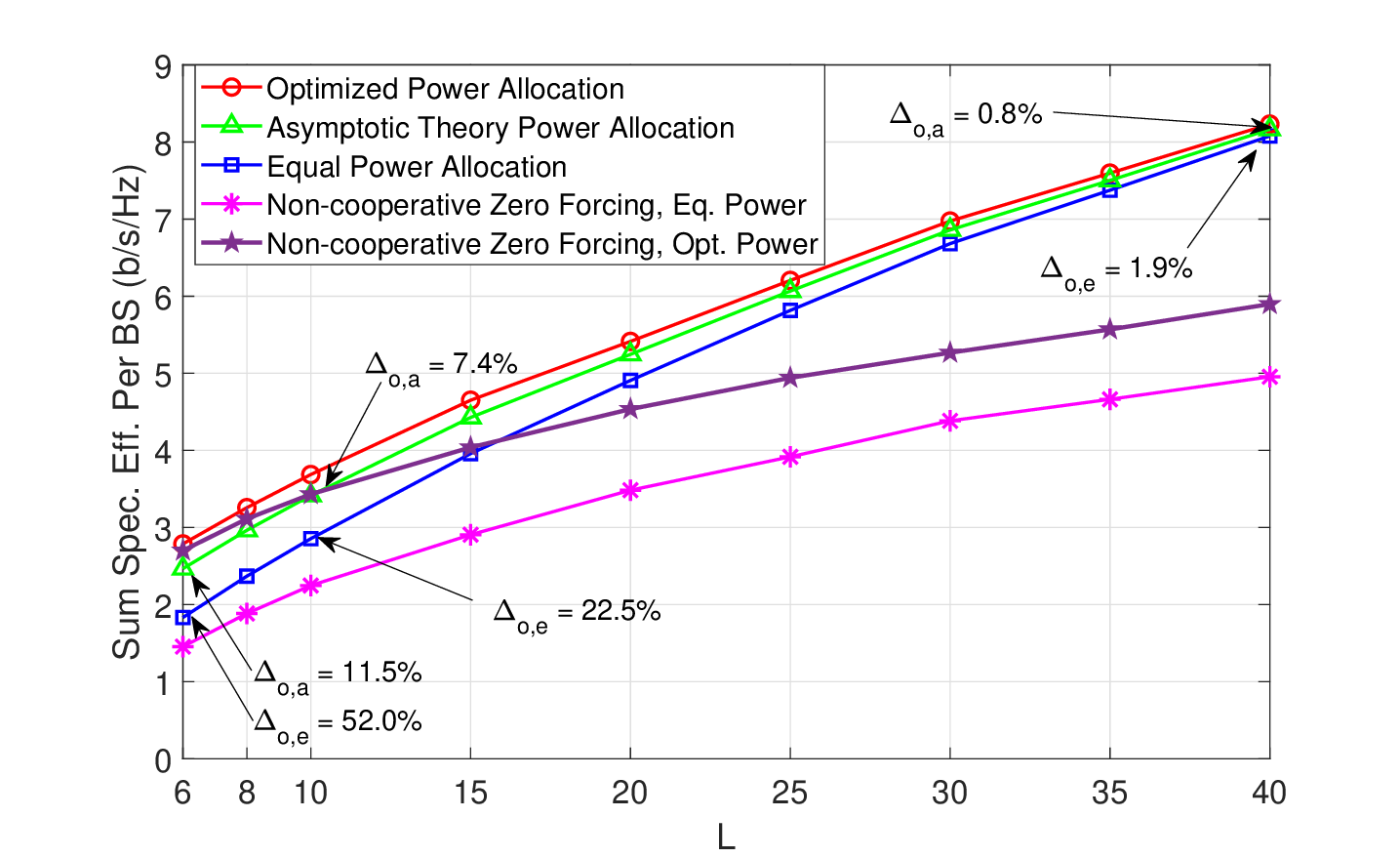}
    \caption{Sum spectral efficiency per base station vs. $L$, for $M=3$ and different power allocation approaches. $\Delta_{o,e}$, and $\Delta_{o,a}$ denote  the percentage difference between the optimized and equal power allocations, and optimized and asymptotic power allocations, respectively.}
    \label{fig:multPowAllocM3}
\end{figure}
Fig. \ref{fig:multPowAllocM3} shows the sum spectral efficiency per base station vs. the number of antennas, $L$, for the three power allocation policies considered. We used a base station density of 60 BS/$km^2$, $\alpha = 3$, $M=3$ active mobiles per cell, cell radius of $0.15 km$ and a constant noise power such that the cell-edge SNR ($PR^{-\alpha}/\sigma^2$) is  $25dB$. Note that  the WMMSE approach has the highest sum spectral efficiency. At $L = 10$, the WMMSE power allocation achieves a throughput that is 29.4\% higher than the throughput of the constant power allocation, but only 7.4\% higher compared to the suggested power allocation (based on Theorem \ref{T:water_filling}). Thus, our novel power allocation is useful even for quite a small number of antennas, leading to throughput that is just a few percent below optimal, while requiring much less computations. For larger numbers of antennas, this difference diminishes. E.g. at $L = 40$ antennas per BS, the discrepancy between the W-MMSE and our approach reduces to 0.8\%. 

\change{For reference, Fig. \ref{fig:multPowAllocM3} also depicts the sum spectral efficiency with non-cooperative zero forcing, where each BS performs zero forcing only to users within its cell. In particular, the figure shows the performance with optimal W-MMSE power control and with equal power allocation.
Note that the non-cooperative zero-forcing schemes perform considerably worse than the cooperative schemes as the number of antennas grows large. For a smaller number of antennas, the non-cooperative zero forcing system with W-MMSE power control is competitive with the cooperative scheme. This shows that optimal power allocation is especially important in non-cooperating networks. Yet, it is no match for BS cooperation with a large number of antennas.}


These results indicate that the simple power allocation policy of Theorem \ref{T:water_filling} results in a  sum spectral efficiency that approaches that of a much more computationally intensive power allocation policy, which requires the power allocations to be optimized across all base stations. While our scheme was derived using asymptotic limits, it is shown to be efficient even when the number of antennas per BS is only moderately large (e.g., 10 or more).  Finally, we observe from Figure \ref{fig:multPowAllocM3} 
that the power allocation policy of Theorem \ref{T:water_filling}  does result in a substantial spectral efficiency increase over an equal power allocation policy for a small number of antennas. This gain comes for only a small increase in complexity (as most computations can be performed locally using equations that only depend on system parameters and in-cell path-losses).


\section{Summary and Conclusions}

We have developed a method to analyze the SINR and spectral efficiency on the downlink of cooperative cellular networks with multi-antenna base stations. The base stations perform partial zero forcing, nulling the interference  to mobiles within a distance $D$ of them. 
This system is complicated to analyze due to the  dependence between the beam-forming vectors used by nearby base stations 
since base stations 
also null users of other cells.  Such dependence between nodes in spatial point processes is known to be challenging to analyze.

To handle this complexity, we derived novel bounds on marked Poisson point processes with  dependent marks. Combining these bounds with an asymptotic analysis, we found a simple closed-form expression for  an appropriately normalized version of the SINR in the limit as the number of antennas per base station grows large. This result is further used to derive asymptotic expressions for the spectral efficiency on the downlink. 

Further, these simple asymptotic expressions for the spectral efficiency are used to derive a simple power allocation scheme which is optimal in the sense of maximizing the mean spectral efficiency of a BS, if the asymptotic expressions are taken as accurate.  This simple power allocation algorithm mostly requires base stations to perform power allocation locally, requiring minimal coordination between base stations on power allocation. The findings of this work are validated by Monte Carlo simulations which show that the asymptotic expressions are useful even when the number of antennas per base station is moderately large. 

 The asymptotic expressions we derived are simple and reveal the dependence of the spectral efficiency on tangible system parameters, such as numbers of antennas per base station, base station density, cell size and mobile distribution. Therefore, the results of this work are useful for system design and system optimization, and also allow better understanding of the network.

\begin{appendices}

\section{Proof of Theorem \ref{th:SN_bounds}} \label{Sec:ProofOfCrossExpectation}
Consider a finite (large) radius $R$, and define $I_k(R)=\sum_i f_k(r_i)p_{i,k} 1_{[r_i<R]}$. Also, consider a division of the area to rings of width $\Delta=R/N$ and define an indicator function $u_n(r)=1_{[(n-1)\Delta \le r <n\Delta} $. Let $q_k[n]=\sup_{(n-1)\Delta \le r< n\Delta} q_k(r)$ (or $q_k[n]=\inf_{(n-1)\Delta \le r< n\Delta} q_k(r)$ for an upper bound), and similarly, $f_k[n]=\sup_{(n-1)\Delta \le r< n\Delta} f_k(r)$ ($f_k[n]=\inf_{(n-1)\Delta \le r< n\Delta} f_k(r)$). We thus have:
\begin{align}
    &E\left[I_1(R)I_2(R)\right]
    \\\notag&=\sum_{n_1=1}^N\sum_{n_2=1}^N E\Big[\sum_{i,j} f_1(r_i) p_{i,1} u_{n_1}( r_{i}) f_2(r_j) p_{j,2} u_{n_2}(r_{j})\Big]
\end{align}
Next, we apply the bounds in \eqref{Eqn:JointBounds} which give:
\begin{align}
    &\sum_{i,j}E\Big[ f_1(r_i) p_{i,1} u_{n_1}( r_{i}) f_2(r_j) p_{j,2} u_{n_2}(r_{j})|r_i,r_j\Big]
    \\\notag&\le(\ge)
    \sum_{i}E\Big[\sum_{i,j} f_1(r_i) f_2(r_i) q_0(r_i) u_{n_1}( r_{i})  |r_i,r_j\Big] \\\notag&+\!\sum_{i\ne j}\!E\Big[ f_1(r_i) f_2(r_j) u_{n_1}( r_{i})   u_{n_2}(r_{j}) q_1\!(r_i)q_2(r_j)|r_i,\!r_j\Big]
\end{align}
Applying the chain rule for expectations and using the quantized bounds  $q_k(r) \le (\ge) q_k[n]$ and $f_k(r) \le (\ge) f_k[n]$ for $(n-1)\Delta \leq r < n\Delta $ results in:
\begin{align}\label{e:temp_bound}
    E\left[I_1(R)I_2(R)\right] 
    &\le (\ge)
    \notag\\&\sum_{n_1=1}^N f_1[n_1] f_2[n_1]q_{0}[n_1] E\Big[ \sum_i u_{n_1}(r_{i}) \Big]
    \notag\\
    &+\sum_{n_1=1}^N\sum_{n_2=1}^N f_1[n_1]  f_2[n_2]q_{1}[n_1]q_{2}[n_2]
    \notag\\&\cdot E\Big[\sum_{i\ne j} u_{n_1}( r_{i})  u_{n_2}( r_{j})\Big]
\end{align}

Since $\sum_i u_{n}(r_{i})$, is a Poisson random variable with parameter 
$\lambda_n^N=\int_{(n-1)\Delta}^{n\Delta} \int_0^{2\pi }r\lambda(r,\theta)d\theta dr$,  $E\big[ \sum_i u_{n_1}(r_{i}) \big]=\lambda_{n_1}^N$ and $E\big[\sum_{i\ne j} u_{n_1}( r_{i})  u_{n_2}( r_{j})\big]=E\big[\sum_{i, j} u_{n_1}( r_{i})  u_{n_2}( r_{j})\big]-E\big[\sum_{i} u_{n_1}( r_{i})  u_{n_2}( r_j)\big]
 =\lambda_{n_1}^N\lambda_{n_2}^N$. Substituting in \eqref{e:temp_bound} we have:
\begin{align}
&E\left[I_1(R)I_2(R)\right]  
     \le (\ge)
         \notag\\&=\sum_{n_1=1}^N\sum_{n_2=1}^N f_1[n_1] q_{1}[n_1] f_2[n_2]q_{2}[n_2]\lambda_{n_1}^N\lambda_{n_2}^N \notag\\&+\sum_{n_1=1}^N f_1[n_1] f_2[n_1]q_{0}[n_1]  \lambda_{n_1}^N
\end{align}

To conclude, we note that for any continuous function, $g(r)$, and any family of sample sequences  $r_{n,N}$ satisfying $(n-1)\frac{R}{N}\le r_{n,N} <n\frac{R}{N}$, we have 
\begin{align}
\lim_{N\to\infty}\sum_{n=1}^N g(r_{n,N})\lambda_n^N&=\int_{0}^{R} \int_0^{2\pi }g(r)r\lambda(r,\theta)d\theta dr
\notag\\&=2\pi\int_{0}^{R} g(r)r\bar \lambda(r) dr. \label{Eqn:SumToIntegral}
\end{align}
Thus,
\begin{align}
E\left[I_1(R)I_2(R)\right]  
     &\le (\ge) 
     \prod_{k=1}^2 \Big(2\pi\int_{0}^{R} f_k(r) q_k(r) r\bar \lambda(r) dr\Big)
      \notag\\&+2\pi\int_{0}^{R} f_1(r) f_2(r) q_0(r) r\bar \lambda(r) dr.\notag
\end{align}
Taking the limit as $R\to\infty$ completes the proof.
\section{Proofs of Lemmas}
\subsection{Proof of Lemma \ref{lemma:InterfConv}} \label{sec:InterfPowerConv}
 Since for large enough $L$, $D>R$,  without loss of generality, we only need to consider the  $D>R$ case. The interference power at the typical mobile is given by 
\begin{align}
 I_0 &
 = \sum_{k = 1}^\infty  a_k\sum_{i\in \mathcal{M}_k}\left|\varphi_{i,k} \mathbf{h}_{0,k}^\dagger\mathbf{w}_{i}\right|^2 = \sum_{k = 1}^\infty 1_{\left\{r_{0, k} > D\right\}}r_{0,k}^{-\alpha}\tilde{a}_k. 
\notag
\end{align}
where 
\begin{align}
 \tilde a_k&=a_k \sum_{i\in \mathcal{M}_k}\left|\varphi_{i,k} r_{0,k}^{\alpha} \mathbf{h}_{0,k}^\dagger\mathbf{w}_{i}\right|^2 
 = a_k \sum_{i\in \mathcal{M}_k}\left|\varphi_{i,k}  \zeta_{i,k}\right|^2  .
\notag
\end{align}
is the effective transmitted power of BS $k$ to the typical mobile, and $\zeta_{i,k}=\mathbf{g}_{0,k}^\dagger\mathbf{w}_{i}$. Since BSs that are further than $D$ from the origin do not consider the typical mobile in their beamforming design, their  precoding vectors, $\mathbf{w}_{i}$ 
$\mathbf{g}_{0,k}$ are independent. Recalling also that $\mathbf{g}_{0,k}\sim\mathcal{CN}(0,\mathbf{I})$ and $\|\mathbf{w}_{i}\|=1$ we have that  $\zeta_{i,k}\sim\mathcal{CN}(0,1)$ and are  i.i.d. random variables for all BSs outside of a disk of radius $D$ from the origin. For $r_{0,k}\le D$, $\tilde a_k$ is always multiplied by zero. Thus, we can assume that $\zeta_{i,k}\sim\mathcal{CN}(0,1)$ and  i.i.d. for any $k>0$.



To show convergence of the normalized interference in the mean-square sense, we first find the limit of the expectation and then show that its variance goes to zero.

\subsubsection{Expected Interference}
From Prop. 2.4 in \cite{baccelli2010}: 
\begin{align}\label{e:EI_def}
    E[I_0] &= 2\pi\int_D^\infty r^{1-\alpha}E[\tilde a_k|r_{0,k}=r]\lambda_b dr.
\end{align}
To bound the expectation, we need upper and lower bounds (for $k > 0$ and $r_{0,k} > D$) on  
\begin{align}\label{e:Expectation_of_tilde_ak}
    E[ \tilde{a}_k|r_{0,k}=r] = \Pr(a_k=1|r_{0,k}=r) \cdot \bar P(r)
\end{align}
where we used the fact that $a_k\in\{0,1\}$ and $\bar P(r)$ was defined in \eqref{e:average_p_vs_distance}. We also define:
\begin{align}
\bar a(r) = \Pr(a_k=1|r_{0,k}=r)  =   E\left[  a_k |r_{0,k} =r\right].
\end{align}
Recall that $a_k  = 1$ if the number of mobiles in a distance less than $D$ from BS $k$ is at most $L$. As we have $M$ mobiles around each BS,  $a_k  = 1$ if  (but not only if) there are less than or equal to $L/M$ base stations closer than distance $D+R$ to BS $k$.  Since we condition on an active mobile being at the origin, there must be one base station at a distance at most $R$ from the origin. This BS should be treated separately from the others. As a worst case assumption, we will derive a bound by assuming that BS $0$ is at a distance of less than $D+R$ from BS $k$. All other BSs are from a HPPP with intensity $\lambda$. Thus, the number of BSs at a distance of at most $D+R$ from BS $k$ is a Poisson random variable with parameter $\pi \lambda_b (D+R)^2$, and the probability that  it  is below $L/M-1$ is
\begin{align}
&Q(\floor{L/M}, \pi \lambda_b (D+R)^2) 
\notag\\&\quad\quad= e^{-\pi \lambda_b (D+R)^2}\sum_{i=0}^{\floor{L/M}-1} \frac{(\pi \lambda_b (D+R)^2)^i}{i!}  \label{Eqn:LowerIncompleteGammaPoissonCDFIdentity} 
\end{align}
where $Q(a, z)=\Gamma(a,z)/\Gamma(a)$. 
Thus
\begin{align}
\bar a(r) \geq Q\left(\left\lfloor{L}/{M}\right\rfloor, \pi \lambda_b (D+R)^2\right)
\label{Eqn:BSFocusedLBak}
\end{align}

Similarly, $a_k = 1$ only if (but not necessarily if) there are less than $L/M$ base stations  closer than $D-R$ to BS $k$. For this bound we assume that BS $0$ is far from BS $k$. Using again the upper regularized gamma function to bound the probability that the number of BSs within a distance $D-R$ from BS $k$ is  smaller than $L/M$ we get:
\begin{align}\label{e:a_k_upper_bound}
\bar a(r) \leq Q\left(\left\lfloor\frac{L}{M}\right\rfloor+1, \pi \lambda_b (D-R)^2\right).
\end{align}
As we assume that $\bar P(r)=\bar P + \xi(r)$ where $|\xi(r)|\le \delta r^{-\gamma}$ with $\gamma>0$, 
using \eqref{e:EI_def}, we have
\begin{align}\label{e:EI_lower_bound}
    E[I] \geq& 2\pi\int_D^\infty r^{1-\alpha}Q\left(\left\lfloor\frac{L}{M}\right\rfloor, \pi \lambda_b (D+R)^2\right) \notag\\&\cdot (\bar P+\xi(r)) \lambda_b dr
    \notag\\
\ge& 2\pi\lambda_b  Q\left(\left\lfloor\frac{L}{M}\right\rfloor, \pi \lambda_b (D+R)^2\right) \frac{D^{2-\alpha}}{\alpha-2}
\notag\\&\cdot\left(\bar P-\delta D^{-\gamma}\frac{\alpha-2}{\alpha+\gamma-2}\right)
\end{align}
and on the other hand, using  \eqref{e:a_k_upper_bound} we get:
\begin{align}\label{e:EI_upper_bound}
    E[I] \leq &2\pi\int_D^\infty r^{1-\alpha}Q\left(\left\lfloor\frac{L}{M}\right\rfloor+1, \pi \lambda_b (D+R)^2\right) \notag\\&\cdot
    (\bar P+\xi(r))\lambda_b dr
    \notag\\
  \leq& 2\pi\lambda_b  Q\left(\left\lfloor\frac{L}{M}\right\rfloor+1, \pi \lambda_b (D+R)^2\right) \frac{D^{2-\alpha}}{\alpha-2}
  \notag\\&\cdot
    \left(\bar P+\delta D^{-\gamma}\frac{\alpha-2}{\alpha+\gamma-2}\right).
\end{align}
Note that both in \eqref{e:EI_lower_bound} and \eqref{e:EI_upper_bound} the second term in the parentheses $\to 0$ as $D\to\infty$.

We next consider the behavior of these expectations at the limit when $L\to\infty$ and $D=s\cdot L^\beta$.
Lemma 2 of \cite{zhu2014performance}, shows that 
$
    \lim_{L\to\infty} Q(L, qL) 
$ converges to $1$ if  $0\le q < 1$ and converges to $0$ if $q\ge 1$. Additionally, since $Q(L, x)$ monotonically decreases with $x$, this is easily extended to show that $
    \lim_{L\to\infty} Q(L, qL^{2\beta}) 
$ converges to $1$ if $\beta<0.5$ and converges to $0$ if $\beta>0.5$.
  Hence, for any $s$ and $B$ combination satisfying the assumptions of the lemma 
the bounds in \eqref{Eqn:BSFocusedLBak} and \eqref{e:a_k_upper_bound} become  
\begin{align}
    \label{e:ar_limit_1}
1\le\lim_{L\rightarrow\infty} \bar a(r)\le 1
.
\end{align}
substituting in \eqref{e:EI_lower_bound} and \eqref{e:EI_upper_bound}, we conclude that 
\begin{align}
    \lim_{L\to\infty} E[D^{\alpha-2}I] &= \frac{2\pi\bar P\lambda_b}{\alpha-2}   \label{Eqn:BSFocusScaledUBLimit}.
\end{align}

\subsubsection{Second Moment of the Interference}
Next, to evaluate the variance of the interference, $I_0$, we bound the second moment of $I_0$ using Theorem \ref{th:SN_bounds}.
 We need to bound  $E[\tilde{a}_k\tilde{a}_\ell|r_{0,k}, r_{0,\ell}]$ for  $k > 0$ and $\ell >0$.
Recalling that $a_k\in\{0,1\}$ we start with:
\begin{align}
    &E[\tilde{a}_k\tilde{a}_\ell|r_{0,k}, r_{0,\ell}] 
    \leq E[\tilde{a}_k\tilde{a}_\ell|r_{0,k}, r_{0,\ell}
,a_k=a_\ell=1]
\label{eqn:InterfCrossBoundGeneral}\notag
\end{align}

For the case that  $k\neq \ell$, we use the fact that $\zeta_{i,k}$  are i.i.d. $\mathcal{CN}(0,1)$ random variables. Thus
\begin{align}
    &E[\tilde{a}_k\tilde{a}_\ell|r_{0,k}, r_{0,\ell}]   
    \\&\notag\quad\leq E\left[\left. \sum_{i\in \mathcal{M}_k}\varphi_{i,k}^2    \sum_{i\in \mathcal{M}_\ell}\varphi_{i,\ell}^2\right|r_{0,k}, r_{0,\ell} ,a_k=a_\ell=1\right]. 
\end{align}
This expectation is bounded by the following lemma.
\begin{lemma}\label{lem:power_correlation}
The correlation between the transmitted power of two base stations, given their distances from the origin is upper bounded by
\begin{align}
    &E\left[\sum_{i\in \mathcal{M}_k}\varphi_{i,k}^2 \sum_{j\in \mathcal{M}_\ell}\varphi_{j,\ell}^2\Big| r_{0,k},r_{0,\ell},a\quad_k=a_\ell=1 \right] 
    \notag\\&\quad\le (\bar P +c r_{0,k}^{-\tilde\gamma/4})(\bar P +cr_{0,\ell}^{-\tilde\gamma/4}) \label{Eqn:CorrBound}
\end{align}
where $\tilde \gamma=\min\{\gamma,0.5\}$ and $c^2=\delta+(1-\frac{\pi^2}{12})^{-1/2} P^2/\pi$.
\end{lemma}
\begin{IEEEproof}
See Appendix \ref{app:proof_of_power_distance_lemma}
\end{IEEEproof}

In  the case that $k = \ell$,
\begin{align}
    &E[\tilde{a}_k^2|r_{0,k}] \leq E\Bigg[\Big(\sum_{i\in \mathcal{M}_k}\varphi_{i,k}^2 \left|\zeta_{i,k}\right|^2 \Big)^2 \Big|r_{0,k} \Bigg]  \\
    &\quad= E\Big[\sum_{i\in \mathcal{M}_k}\varphi_{i,k}^4\left|\zeta_{i,k}\right|^4  
    \notag\\&\quad \ + \sum_{i\in \mathcal{M}_k}\sum_{j\in \mathcal{M}_k, j\neq i}\varphi_{i,k}^2 \varphi_{j,k}^2\big|\zeta_{i,k}\big|^2\left|\zeta_{j,k}\right|^2 \Big|r_{0,k}\Big]     \notag \\
&\quad= E\left[\left.\sum_{i\in \mathcal{M}_k} \varphi_{i,k}^4  + \sum_{i\in \mathcal{M}_k}\sum_{j\in \mathcal{M}_k}\varphi_{i,k}^2 \varphi_{j,k}^2 \right|r_{0,k} \right] 
\notag\end{align}
The power constraint, $\sum_{i\in \mathcal{M}_k}\varphi_{i,k}^2\le P$, also induces $\sum_{i\in \mathcal{M}_k}\varphi_{i,k}^4\le P^2$, and hence we have:
\begin{align}
    E[\tilde{a}_k^2|r_{0,k}]\le  2 \brP^2 \label{eqn:InterfCrossBoundEqualTerms}.
\end{align}

Using Theorem \ref{th:SN_bounds} with $f_1(r)=f_2(r)=r^{-\alpha} \cdot 1_{\{r > D\}}$, $p_{1,k}=p_{2,k}=\tilde a_k$,  
    $q_0(r) = 2\brP^2 \cdot $ and
    $q_1(r) =q_2(r) = (\bar P +cr^{-\tilde\gamma/4}) $, yields
\begin{align}
    E\left[ I_0^2 \right] \leq & \left(2\pi \lambda_b\int_D^\infty (\bar P +cr^{-\tilde\gamma/4}) r^{-\alpha+1} dr \right)^2 
    \notag\\&+ 4\pi \brP^2\lambda_b\int_{D}^\infty r^{-2\alpha + 1} dr  
\end{align}
Solving the integral and using also \eqref{e:EI_lower_bound} we have
\begin{align}
    &\var{I_0}=E[I_0^2]-E^2[I_0]
     \\ &\le \left(\frac{2\pi \bar P \lambda_b D^{2-\alpha}}{\alpha-2}\right)^2\left( 1 +\frac{ c (\alpha-2) D^{-\tilde \gamma/4}}{\bar P(\alpha+\tilde\gamma/4-2)}\right)^2 
    \notag\\
    &+  \frac{4\pi \brP^2\lambda_b D^{2-2\alpha}}{2\alpha -2}     -Q^2\left(\floor{\frac{L}{M}}+1, \pi \lambda_b (D+R)^2\right) \notag\\&\cdot\left(\frac{2\pi\lambda_b D^{2-\alpha}}{\alpha-2}\right)^2\left(\bar P-\delta D^{-\gamma}\frac{\alpha-2}{\alpha+\gamma-2}\right)^2.
    \notag
\end{align}
With the nulling radius $D$ such that  $s>0$ with $0<\beta<0.5$ or $0<s<1$ with $\beta=0.5$, the limits in \eqref{e:ar_limit_1} and the fact that $\alpha > 2$, we get for any $\gamma>0$ that  $\var{D^{\alpha-2}I_0}$ scales as $ O(D^{4-2\alpha})$. So $\lim_{D\to\infty} \var{D^{\alpha-2}I_0}=0$ which 
combined with \eqref{Eqn:BSFocusScaledUBLimit} completes the proof. 
\subsection{Proof of Lemma \ref{Lemma:SignalPower}} \label{Lemma:SignalPowerProof}
$\mathbf{Q}_{0,0}$ is a projection matrix and thus is idempotent, and is also Hermitian. Hence,   $\|\mathbf{Q}_{0,0}\mathbf{g}_{0,0}\|=\sqrt{\mathbf{g}_{0,0}^\dagger\mathbf{Q}_{0,0}\mathbf{g}_{0,0}}$. Substituting in \eqref{eqn:S0def} and using also  $a_0\in\{0,1\}$ gives
\begin{align}
    S_0= \varphi_{0,0}^2a_0r^{-\alpha}_{0,0} \mathbf{g}^\dagger_{0, 0} \, \mathbf{Q}_{0,0} \mathbf{g}_{0,0} \label{Eqn:SignalPower}
\end{align}

Note that given the number of zero-forced mobiles, $\sum_{j} 1_{r_{j,0}< D}$, the quantity $2\cdot \mathbf{g}^\dagger_{0, 0}\mathbf{Q}_{0,0} \mathbf{g}_{0,0}$
 is a $\chi^2$ random variable with
$2(L-\sum_{j} 1_{r_{j,0}< D})$ degrees of freedom.
Hence, $ \mathbf{g}^\dagger_{0, 0}\mathbf{Q}_{0,0} \mathbf{g}_{0,0}$   is the sum of $(L-\sum_{j} 1_{r_{j,0}< D})$, unit mean, exponential random variables. Recalling that  $D = sL^\beta$, with $0 < \beta \leq 0.5$,  from the strong law of large numbers we have with probability 1 that:
\begin{align}
    \lim_{L\to\infty}\frac{\mathbf{g}^\dagger_{0, 0}\mathbf{Q}_{0,0} \mathbf{g}_{0,0}}{L  - \sum_{j} 1_{r_{j,0}< D}} = 1.
    \label{e:num_DOF_D}
\end{align}

As the BSs form a HPPP, and there are  $M$ mobiles per  base station,   with probability 1,
\begin{align}
    \lim_{D\to\infty} \frac{1}{ D^2} \sum_{j} 1_{r_{j,0}< D} = \pi\lambda_b M \label{Eqn:LimitOfNormalizedNulledMobiles}
\end{align}
Combining with \eqref{e:num_DOF_D}:
\begin{align}
    \lim_{L\to\infty}\frac{\mathbf{g}^\dagger_{0, 0}\mathbf{Q}_{0,0} \mathbf{g}_{0,0}}{L  - s^2L^{2\beta}\pi\lambda_b M}  = 1  \label{Eqn:ChiSquareLimit}
\end{align}
Substituting into \eqref{Eqn:SignalPower},
 and recalling that $a_0\to 1$  as $L\to \infty$ in MSE  completes the proof.

\subsection{Proof of Lemma \ref{lem:power_correlation}}\label{app:proof_of_power_distance_lemma}
For any two BSs, we assumed that 
\begin{align}
    C_{k,\ell}(\Phi_b)&\triangleq E\left[\sum_{i\in \mathcal{M}_k}\varphi_{i,k}^2 \sum_{j\in \mathcal{M}_\ell}\varphi_{j,\ell}^2\Big|\Phi_b,a_k=a_\ell=1\right]\notag\\& = \bar P^2 + \xi_{k,\ell}
\end{align}
with $|\xi_{k,\ell}|\le \min\{P^2,\delta \tilde r_{k,\ell}^{-\gamma}\}$. Removing the condition on all BS locations and conditioning only on $r_{0,k}$, $r_{0,\ell}$:
%
%
\begin{align}\label{e:was 81}
    E&\left[\sum_{i\in \mathcal{M}_k}\varphi_{i,k}^2 \sum_{j\in \mathcal{M}_\ell}\varphi_{j,\ell}^2\Big| r_{0,k},r_{0,\ell},a_k=a_\ell=1 \right] 
    \notag\\
    &= E\left[E\left[C_{k,\ell}(\Phi_b) \right]\Big| r_{0,k},r_{0,\ell},a_k=a_\ell=1\right]
    \notag\\&
    \le \bar P^2 +E\left[  |\xi_{k,\ell}|\Big| r_{0,k},r_{0,\ell}\right].
\end{align}
Let $q=\sqrt{r_{0,k}r_{0,\ell}}.$ We have:
\begin{align}\label{e:E_zeta}
    E&\left[  |\xi_{k,\ell}|\Big| r_{0,k},r_{0,\ell}\right]
    \notag\\&=\Pr(\tilde r_{k,\ell}\le q)E\left[  |\xi_{k,\ell}|\Big| r_{0,k},r_{0,\ell},\tilde r_{k,\ell}\le q\right]
    \notag\\&+\Pr(\tilde r_{k,\ell}> q)E\left[  |\xi_{k,\ell}|\Big| r_{0,k},r_{0,\ell},\tilde r_{k,\ell}> q\right]
    \notag\\&
    \le\Pr(\tilde r_{k,\ell}\le q)  P^2+\delta q^{-\gamma}.
\end{align}

Let $\theta$ be the angle between the BSs as seen by the typical mobile, and using the law of cosines:
\begin{align}\label{e:law_of_cosines}
    \tilde r_{k,\ell}= r_{0,k}^2+r_{0,\ell}^2-2r_{0,k}r_{0,\ell}cos\theta.
\end{align}    
Using \eqref{e:law_of_cosines}, the probability that the distance between the two BSs is smaller than $q$ is 
\begin{align}
    \Pr(\tilde r_{k,\ell}\le q)&=\Pr\left(cos\theta\ge\frac{r_{0,k}^2+r_{0,\ell}^2-\sqrt{r_{0,k}r_{0,\ell}}}{2r_{0,k}r_{0,\ell}} \right)
    \notag\\&\le
    \Pr\left(1-cos\theta\le \frac{\sqrt{r_{0,k}r_{0,\ell}}}{2r_{0,k}r_{0,\ell}} \right).
\end{align}
Using also the inequality $1-\cos\theta\ge \frac{\theta^2}{2}-\frac{\theta^4}{24}\ge (1-\frac{\pi^2}{12})\frac{\theta^2}{2} $ for $|\theta|\le \pi$ we have
\begin{align}
 &   \Pr(\tilde r_{k,\ell}\le q)\le \Pr\left((1-\frac{\pi^2}{12})\frac{\theta^2}{2}
    \le \frac{\sqrt{r_{0,k}r_{0,\ell}}}{2r_{0,k}r_{0,\ell}} \right)
    \\\notag&\;\;\;\;\;\;=\Pr\left(|\theta|\le (1-\frac{\pi^2}{12})^{-1/2}(r_{0,k}r_{0,\ell})^{-1/4} \right).
\end{align}
Finally, since the network is homogenous, $\theta$ is uniformly distributed over $[0,2\pi)$. Thus
\begin{align}
    \Pr(\tilde r_{k,\ell}\le q)\le\frac{1}{\pi}{\left(1-\frac{\pi^2}{12}\right)^{-1/2} (r_{0,k}r_{0,\ell})^{-1/4} }.
\end{align}
Substituting in \eqref{e:E_zeta} and then in \eqref{e:was 81}, we have:
\begin{align*}
    &E\left[\sum_{i\in \mathcal{M}_k}\varphi_{i,k}^2 \sum_{j\in \mathcal{M}_\ell}\varphi_{j,\ell}^2\Big| r_{0,k},r_{0,\ell},a_k=a_\ell=1 \right] 
    \notag \\ &\quad\quad\le \bar P^2 +c^2\cdot(r_{0,k}r_{0,\ell})^{-\frac{\tilde\gamma}{2}}
    \notag \\ &\quad\quad\le (\bar P +c r_{0,k}^{-\frac{\tilde\gamma}{4}})(\bar P +cr_{0,\ell}^{-\frac{\tilde\gamma}{4}}).
\end{align*}
where we used the lemma definitions for $c^2$
 and $\tilde \gamma$. 

\section{Proof of Mean-Square Convergence of the Throughput}
\label{Sec:MSEConvergence}

Let $\breve \eta_0={\varphi_{0,0}^2 r^{-\alpha}_{0,0} \mathbf{g}^\dagger_{0, 0}  \mathbf{g}_{0,0}}/({  \mu L^{1-\alpha/2}})$,  $H~=~\lim_{L\to\infty}L^{-\alpha/2}\eta_0=\frac{(1 - \tilde s^2\pi\lambda_b M)\varphi_{0,0}^2r^{-\alpha}_{0,0}}{\frac{1}{s}\frac{2\pi\bar P\lambda_b}{\alpha-2} + \mu }$ and $\breve H = \lim_{L\to\infty} L^{-\alpha/2}\breve \eta_0  = {\varphi_{0,0}^2 r^{-\alpha}_{0,0}}/{\mu}$.
 As $\mathbf{g}^\dagger_{0, 0}  \mathbf{g}_{0,0}$ is a $\chi ^2$ random variable with $2L$ degrees of freedom, $L^{-\alpha/2}\breve \eta_0$ has a bounded third moment. Noting that
 $0\le \eta_0\le \breve \eta_0$, both $L^{-\alpha/2}\breve \eta_0$ and $L^{-\alpha/2} \eta_0$ are uniformly integrable random variables  \cite{gut2005probability} and hence
$\lim_{L\to\infty } E\left[\left|L^{-\alpha/2} \eta_0  -  H\right|^2\right] = 0$.

Let $\mathcal Y$ be $1$ if ${ \eta_0\ge L^{\alpha/2} H/4}$, and $0$ otherwise, and consider $\mathcal E=E[|\log_2(1+\eta_0)-\log_2(1+L^{\alpha/2}H)|^2]$. 
We need to prove that $\lim_{L\to\infty}\mathcal E=0$. We have:
\begin{IEEEeqnarray}{rcl}
\label{Eqn:SpecEffBound}
\mathcal E&=&
E[|\log_2(1+\eta_0)-\log_2(1+L^{\alpha/2}H)|^2 \cdot\mathcal{Y}]
\\\notag&&+E[|\log_2(1+\eta_0)-\log_2(1+L^{\alpha/2}H)|^2\cdot \bar{\mathcal Y}]
\\ &\le &
E\left[\left|\log_2\left(\frac{\eta_0}{L^{\alpha/2}H}\right)\right|^2 \mathcal Y\right]
+E\left[\log_2^2(1+L^{\alpha/2}H) \bar{\mathcal Y}\right] \notag
\end{IEEEeqnarray}
The first term converges to zero because ${\eta_0}/{L^{\alpha/2}H}\to 1$ in MSE  and $|\log x|$ is continuous with bounded slope   for $x>1/4$.
The second term is bound by 
$
\log_2^2(1+L^{\alpha/2}H)\cdot\Pr\left\{\mathcal{Y}=0\right\} 
$ where the first term scales as $\log_2^2(L)$ and the second term can be bounded by
\begin{align}\label{e:Pr_bound_short}
&\Pr\left\{\mathcal{Y}=0\right\}
\le \Pr\left\{a_0<1\right\}
\\\notag&+    \Pr\left\{\frac{\mathbf{g}^\dagger_{0, 0}  \mathbf{Q}_{0,0}\mathbf{g}_{0,0}}{L}<\frac{1 - \tilde s^2\pi\lambda_b M}{2} \right\}
    \\&
    +\Pr\left\{\frac{\sum_{i = 1}^\infty\left|a_{b_i}\varphi_{i,b_i} \mathbf{h}_{0,b_i}^\dagger\mathbf{w}_{i}\right|^2+\mu}{2L^{1-\alpha/2}}>\frac{1}{s}\frac{2\pi\bar P\lambda_b}{\alpha-2} + \mu\right\}\notag
\end{align}
Using the results derived above, it can be shown that all three terms in \eqref{e:Pr_bound_short} decay to zero, where the  slowest term (the third) decays as
$ O(L^{2-\alpha})$. This is still much faster than the $\log_2^2 L$ term. 
 Thus \eqref{Eqn:SpecEffBound} converges to zero, which completes the proof.

\end{appendices}

\bibliographystyle{IEEEtran}
\bibliography{IEEEabrv,main}

\newcommand{\noopsort}[1]{} \newcommand{\printfirst}[2]{#1} \newcommand{\singleletter}[1]{#1} \newcommand{\switchargs}[2]{#2#1}
\begin{thebibliography}{10}
\providecommand{\url}[1]{#1}
\csname url@samestyle\endcsname
\providecommand{\newblock}{\relax}
\providecommand{\bibinfo}[2]{#2}
\providecommand{\BIBentrySTDinterwordspacing}{\spaceskip=0pt\relax}
\providecommand{\BIBentryALTinterwordstretchfactor}{4}
\providecommand{\BIBentryALTinterwordspacing}{\spaceskip=\fontdimen2\font plus
\BIBentryALTinterwordstretchfactor\fontdimen3\font minus \fontdimen4\font\relax}
\providecommand{\BIBforeignlanguage}[2]{{%
\expandafter\ifx\csname l@#1\endcsname\relax
\typeout{** WARNING: IEEEtran.bst: No hyphenation pattern has been}%
\typeout{** loaded for the language `#1'. Using the pattern for}%
\typeout{** the default language instead.}%
\else
\language=\csname l@#1\endcsname
\fi
#2}}
\providecommand{\BIBdecl}{\relax}
\BIBdecl

\bibitem{marzetta2010noncooperative}
T.~L. Marzetta, ``Noncooperative cellular wireless with unlimited numbers of base station antennas,'' \emph{{IEEE} Trans. Commun.}, vol.~9, no.~11, pp. 3590--3600, 2010.

\bibitem{rusek2012scaling}
F.~Rusek, D.~Persson, B.~K. Lau, E.~G. Larsson, T.~L. Marzetta, O.~Edfors, and F.~Tufvesson, ``Scaling up mimo: Opportunities and challenges with very large arrays,'' \emph{IEEE signal processing magazine}, vol.~30, no.~1, pp. 40--60, 2012.

\bibitem{marzetta2015massive}
T.~L. Marzetta, ``Massive {MIMO}: an introduction,'' \emph{Bell Labs Technical Journal}, vol.~20, pp. 11--22, 2015.

\bibitem{larsson2014massive}
E.~G. Larsson, O.~Edfors, F.~Tufvesson, and T.~L. Marzetta, ``Massive {MIMO} for next generation wireless systems,'' \emph{{IEEE} Commun. Mag.}, vol.~52, no.~2, pp. 186--195, Feb. 2014.

\bibitem{bjornson2015massive}
E.~Bj{\"o}rnson, M.~Matthaiou, and M.~Debbah, ``Massive {MIMO} with non-ideal arbitrary arrays: Hardware scaling laws and circuit-aware design,'' \emph{{IEEE} Trans. Commun.}, vol.~14, no.~8, pp. 4353--4368, Apr. 2015.

\bibitem{lu2014overview}
L.~Lu, G.~Y. Li, A.~L. Swindlehurst, A.~Ashikhmin, and R.~Zhang, ``An overview of massive {MIMO}: Benefits and challenges,'' \emph{{IEEE} J. Sel. Topics Signal Process.}, vol.~8, no.~5, pp. 742--758, 2014.

\bibitem{marzetta2016fundamentals}
T.~L. Marzetta and H.~Q. Ngo, \emph{Fundamentals of massive MIMO}.\hskip 1em plus 0.5em minus 0.4em\relax Cambridge University Press, 2016.

\bibitem{bjornson2020making}
E.~Bj{\"o}rnson and L.~Sanguinetti, ``Making cell-free massive {MIMO} competitive with {MMSE} processing and centralized implementation,'' \emph{{IEEE} Trans. Wireless Commun.}, vol.~19, no.~1, pp. 77--90, Jan. 2020.

\bibitem{Govindasamy2018Uplink}
S.~Govindasamy and I.~Bergel, ``Uplink performance of multi-antenna cellular networks with co-operative base stations and user-centric clustering,'' \emph{{IEEE} Trans. Wireless Commun.}, vol.~17, no.~4, pp. 2703--2717, Apr. 2018.

\bibitem{hoydis2013massive}
J.~Hoydis, S.~ten Brink, and M.~Debbah, ``Massive {MIMO} in the {UL/DL} of cellular networks: How many antennas do we need?'' \emph{{IEEE} J. Sel. Areas Commun.}, vol.~31, no.~2, pp. 160--171, Feb. 2013.

\bibitem{yang2013performance}
H.~Yang and T.~L. Marzetta, ``Performance of conjugate and zero-forcing beamforming in large-scale antenna systems,'' \emph{{IEEE} J. Sel. Areas Commun.}, vol.~31, no.~2, pp. 172--179, 2013.

\bibitem{jin2015massive}
S.~Jin, X.~Wang, Z.~Li, K.-K. Wong, Y.~Huang, and X.~Tang, ``On massive {MIMO} zero-forcing transceiver using time-shifted pilots,'' \emph{IEEE Transactions on Vehicular Technology}, vol.~65, no.~1, pp. 59--74, 2015.

\bibitem{bjornson2016massive}
E.~Bj{\"o}rnson, E.~G. Larsson, and M.~Debbah, ``Massive mimo for maximal spectral efficiency: How many users and pilots should be allocated?'' \emph{IEEE Transactions on Wireless Communications}, vol.~15, no.~2, pp. 1293--1308, Feb. 2016.

\bibitem{ammar2020statistical}
H.~A. Ammar, R.~Adve, S.~Shahbazpanahi, and G.~Boudreau, ``Statistical analysis of downlink zero-forcing beamforming,'' \emph{{IEEE} Commun. Lett.}, vol.~9, no.~11, pp. 1965--1969, 2020.

\bibitem{kountouris2009transmission}
M.~Kountouris and J.~G. Andrews, ``Transmission capacity scaling of {SDMA} in wireless ad hoc networks,'' in \emph{2009 IEEE Information Theory Workshop}.\hskip 1em plus 0.5em minus 0.4em\relax IEEE, 2009, pp. 534--538.

\bibitem{wang2020uplink}
Z.~Wang, J.~Zhang, E.~Bj{\"o}rnson, and B.~Ai, ``Uplink performance of cell-free massive {MIMO} over spatially correlated {Rician} fading channels,'' \emph{IEEE Communications Letters}, vol.~25, no.~4, pp. 1348--1352, 2020.

\bibitem{interdonato2020local}
G.~Interdonato, M.~Karlsson, E.~Bj{\"o}rnson, and E.~G. Larsson, ``Local partial zero-forcing precoding for cell-free massive mimo,'' \emph{IEEE Transactions on Wireless Communications}, vol.~19, no.~7, pp. 4758--4774, 2020.

\bibitem{george2017ergodic}
G.~George, R.~K. Mungara, A.~Lozano, and M.~Haenggi, ``Ergodic spectral efficiency in {MIMO} cellular networks,'' \emph{{IEEE} Trans. Wireless Commun.}, vol.~16, no.~5, pp. 2835--2849, Mar. 2017.

\bibitem{dhillon2012modeling}
H.~S. Dhillon, R.~K. Ganti, F.~Baccelli, and J.~G. Andrews, ``Modeling and analysis of k-tier downlink heterogeneous cellular networks,'' \emph{{IEEE} J. Sel. Areas Commun.}, vol.~30, no.~3, pp. 550--560, 2012.

\bibitem{AnsumanAdhikary_2013}
A.~Adhikary, J.~Nam, J.-Y. Ahn, and G.~Caire, ``Joint spatial division and multiplexing—the large-scale array regime,'' \emph{{IEEE} Trans. Inf. Theory}, vol.~59, no.~10, pp. 6441--6463, 2013.

\bibitem{Yin_2013}
H.~Yin, D.~Gesbert, M.~Filippou, and Y.~Liu, ``A coordinated approach to channel estimation in large-scale multiple-antenna systems,'' \emph{{IEEE} J. Sel. Areas Commun.}, vol.~31, no.~2, pp. 264--273, 2013.

\bibitem{Bjornson_2017}
E.~Bj{\"o}rnson, J.~Hoydis, L.~Sanguinetti \emph{et~al.}, ``Massive mimo networks: Spectral, energy, and hardware efficiency,'' \emph{Foundations and Trends{\textregistered} in Signal Processing}, vol.~11, no. 3-4, pp. 154--655, 2017.

\bibitem{ghazanfari2021model}
A.~Ghazanfari, T.~Van~Chien, E.~Bj{\"o}rnson, and E.~G. Larsson, ``Model-based and data-driven approaches for downlink massive {MIMO} channel estimation,'' \emph{{IEEE} Trans. Commun.}, vol.~70, no.~3, pp. 2085--2101, 2021.

\bibitem{banerjee2022downlink}
B.~Banerjee, R.~C. Elliott, W.~A. Krzymie{\'n}, and H.~Farmanbar, ``Downlink channel estimation for {FDD} massive mimo using conditional generative adversarial networks,'' \emph{{IEEE} Trans. Wireless Commun.}, 2022.

\bibitem{wang2022uplink}
Z.~Wang, J.~Zhang, B.~Ai, C.~Yuen, and M.~Debbah, ``Uplink performance of cell-free massive {MIMO} with multi-antenna users over jointly-correlated {Rayleigh} fading channels,'' \emph{{IEEE} Trans. Wireless Commun.}, vol.~21, no.~9, pp. 7391--7406, 2022.

\bibitem{Stoyan}
D.~Stoyan, W.~S. Kendall, and J.~Mecke, \emph{Stochastic Geometry and Its Applications}.\hskip 1em plus 0.5em minus 0.4em\relax John Wiley and Sons, 1995.

\bibitem{haenggi2012stochastic}
M.~Haenggi, \emph{Stochastic geometry for wireless networks}.\hskip 1em plus 0.5em minus 0.4em\relax Cambridge University Press, 2012.

\bibitem{Zhu2018Stochastic}
C.~Zhu and W.~Yu, ``Stochastic modeling and analysis of user-centric network {MIMO} systems,'' \emph{{IEEE} Trans. Commun.}, vol.~66, no.~12, pp. 1--1, Dec. 2018.

\bibitem{clarke2013functional}
F.~Clarke, \emph{Functional analysis, calculus of variations and optimal control}, ser. Graduate Texts in Mathematics, 2013, vol. 264.

\bibitem{chakraborty2020efficient}
S.~Chakraborty, {\"O}.~T. Demir, E.~Bj{\"o}rnson, and P.~Giselsson, ``Efficient downlink power allocation algorithms for cell-free massive {MIMO} systems,'' \emph{IEEE Open Journal of the Communications Society}, vol.~2, pp. 168--186, 2020.

\bibitem{baccelli2010}
\BIBentryALTinterwordspacing
F.~Baccelli and B.~Błaszczyszyn, ``Stochastic geometry and wireless networks: Volume i theory,'' \emph{Foundations and Trends® in Networking}, vol.~3, no. 3–4, pp. 249--449, 2010. [Online]. Available: \url{http://dx.doi.org/10.1561/1300000006}
\BIBentrySTDinterwordspacing

\bibitem{zhu2014performance}
J.~Zhu, S.~Govindasamy, and J.~Hwang, ``Performance of multiantenna linear {MMSE} receivers in doubly stochastic networks,'' \emph{{IEEE} Trans. Commun.}, vol.~62, no.~8, pp. 2825--2839, 2014.

\bibitem{gut2005probability}
\BIBentryALTinterwordspacing
A.~Gut, \emph{Probability: A Graduate Course}, ser. Springer Texts in Statistics.\hskip 1em plus 0.5em minus 0.4em\relax Springer New York, 2005. [Online]. Available: \url{https://books.google.com/books?id=CnBxaLaOcBAC}
\BIBentrySTDinterwordspacing

\end{thebibliography}

\end{document}